%% file: main.tex
\documentclass[twocolumn, superscriptaddress]{revtex4-1}

\input{preamble}

\begin{document}

\title{\textbf{Cell and nuclear shapes predict confined motility}}

\author{Malcolm Mehrabian-Terlexis}
\thanks{These authors contributed equally (co-first authors).}
\affiliation{Laboratoire de Physique de l'École Normale Supérieure, ENS, Université PSL, CNRS, Sorbonne Université, Université Paris Cité, F-75005 Paris, France}

\author{Jacqueline Janssen}
\thanks{These authors contributed equally (co-first authors).}
\affiliation{Laboratoire de Physique de l'École Normale Supérieure, ENS, Université PSL, CNRS, Sorbonne Université, Université Paris Cité, F-75005 Paris, France}

\author{Paul Sitoleux}
\thanks{These authors contributed equally (co-first authors).}
\affiliation{Laboratoire de Physique de l'École Normale Supérieure, ENS, Université PSL, CNRS, Sorbonne Université, Université Paris Cité, F-75005 Paris, France}
\affiliation{Laboratoire Jean Perrin, Sorbonne Université, CNRS, F-75005 Paris France}

\author{Thierry Mora}
\thanks{These authors contributed equally (co-last authors).}
\affiliation{Laboratoire de Physique de l'École Normale Supérieure, ENS, Université PSL, CNRS, Sorbonne Université, Université Paris Cité, F-75005 Paris, France}

\author{Aleksandra M. Walczak}
\thanks{These authors contributed equally (co-last authors).}
\affiliation{Laboratoire de Physique de l'École Normale Supérieure, ENS, Université PSL, CNRS, Sorbonne Université, Université Paris Cité, F-75005 Paris, France}

\author{C\'{e}cile Sykes}
\thanks{These authors contributed equally (co-last authors).}
\affiliation{Laboratoire de Physique de l'École Normale Supérieure, ENS, Université PSL, CNRS, Sorbonne Université, Université Paris Cité, F-75005 Paris, France}

\date{November $24^{\mathrm{th}}$ 2026}

\begin{abstract}
Cancer metastasis relies on an enhanced property of cells to migrate efficiently even through constrictions smaller than their nucleus. Both cell and nuclear morphology may influence this process. How these morphologies relate to the likelihood that a cell will transit through a constriction remains unclear. Here we combine microfluidic experiments with machine learning to predict whether MDA-MB-231 breast cancer cells are able to transit through narrow constrictions. Cells migrate within vertically confined boxes, and may migrate into neighboring boxes by transiting through constrictions, requiring a deformation of the nucleus. Defining a set of 12 geometrical features derived from cell and nuclear contours, we find that their average values are not statistically indicative of whether or not a cell will transit. To overcome this limitation, we use trajectories to train classifiers and evaluate their ability to predict future transit events.  We find that the cell area and cell circularity carry most of the predictive information and less circular cells are more likely to transit. Without retraining, the models also predict transit through narrower constrictions and in more adhesive substrates. 
Morphological features of cells about 30 minutes before a transit event proved most informative about future transit events occurring or not. Together, our results show that cell shape, readable from simple contour descriptors, contains a transferable but condition-dependent signature of imminent transit.
\end{abstract}

\maketitle

\section*{Introduction}

The severity of cancer is closely related to metastasis, during which cancer cells migrate toward blood vessels and invade other organs by squeezing through confined spaces~\cite{salminen2020vitro,perea2021squeezing}. This process has been studied extensively in the context of the epithelial to mesenchymal transition, where cells change phenotype from stationary epithelial, to highly motile mesenchymal state~\cite{roche2018epithelial,mittal2018epithelial}. Recent advances in microfluidics allow the use of precisely engineered geometries for the study of single-cell motility in complex environments~\cite{charras2014physical,paul2017cancer} with well controlled geometries \cite{stoberl2024nuclear,davidson2014nuclear,lautscham2015migration,sadjadi2022ameboid}. Furthermore, microfluidic studies indicate that the deformation of the nucleus during invasion may act as a rate limiting step~\cite{mcgregor2016squish,friedl2011nuclear}. Although cellular and nuclear morphologies are 
important variables for the dynamics of invasion, their predictive role for efficient confined migration remains unclear. 

Depending on the local geometry, migrating cells can adopt a variety of motility strategies. The mesenchymal mode of migration involves the formation of actin rich protrusions in the form of lamellipodia or filopodia, which bind to the environment with focal adhesions~\cite{arjonen2011filopodia}. An alternative strategy is amoeboid motion, in which cells weakly adhere to the environment and move using acto-myosin contractility that generates protrusions known as blebs~\cite{guck2010critical,lammermann2009mechanical}.  Individual cancer cells often employ both strategies at once, or alternate between the two~\cite{bergert2012cell,ruprecht2015cortical}. Compared to the mesenchymal phenotype, amoeboid cells are rounder. These modes of migration, based on the dynamic organisation of the cytoskeleton, are very conserved among species, regardless of whether they possess a nucleus (eukaryotic cells) or not (prokaryotic cells)~\cite{radler2026dynamic}. While cell morphology is a good  indicator of the strategy employed for cell migration, the role of the nucleus during cell migration through confined spaces may not be as passive as initially considered. Indeed, the nucleus plays an important role in gene regulation, and also an impact on biomechanics~\cite{friedl2011nuclear,amiri2024inferring,keys2024rear}. As the stiffest organelle in the cell, it must be deformed as the cell moves through confined spaces, directly affecting migration dynamics~\cite{renkawitz2019nuclear,venturini2020nucleus}. Nuclear plasticity may enable highly motile healthy cells such as neutrophils to navigate spaces~\cite{karling2025immune}. Similarly, metastatic cancer cells can have highly deformable nuclei, increasing their migration efficiency under confinement~\cite{van2025deformability}. Therefore, cell and nuclear morphologies are expected to be valuable sources of information pertaining to the state of the cell. They are thus fundamental for a mechanistic understanding of motility under confinement. While several works have used some morphological descriptors for theoretical modeling of cell migration~\cite{godeau20203d,bruckner2022geometry, wortel2021local,moure2017phase}, a full description of the cell and nuclear contour dynamics could be a source of information on the ability of a cell to pass through constrictions. However, such an approach requires a large number of parameters which are difficult to measure experimentally. 

Machine-learning (ML) methods are well suited for handling complex multivariate data for predictive tasks. ML has been used extensively on genomic data for predicting metastasis based on transcriptomics data ~\cite{hirway2023review}, and in vitro assays~\cite{bahar2021modeling,umar2023prediction}. Recently, ML has been used to classify cell migration types based on raw single cell trajectories~\cite{kimmel2019deep,mencattini2020discovering}, and contour-derived features to classify morphology types of cells migrating in extra-cellular matrix ~\cite{eddy2021morphodynamics}. In another work, cell and mitochondria fluorescence images were used to extract features to predict the velocity of cells migrating under 5\um vertical confinement~\cite{zhang2018morphology}. Here, we observe the morphology of motile cells and their nucleus in a microfluidic device made of flat boxes connected with narrow constrictions. A set of features are chosen to describe cell and nuclear morphology using their contours. This data set is then used to train a ML model capable of successfully predicting the transit probability. This model is based on a limited amount of features, and cell circularity, as well as area, are among the most predictive. The model is successfully transferred to other experimental conditions, on the one hand to smaller constrictions and, on the other hand, in enhanced cell adhesion conditions by coating the microfluidic surfaces with fibronectin. We extend the model to predict the time at which a given cell will transit, revealing a time scale of 30 minutes during which the model is most predictive.

\section*{Results}

\subsection*{Trajectories and transits}
\begin{figure*}[tp]
    \centering
        \includegraphics[
            width=\textwidth,
        ]{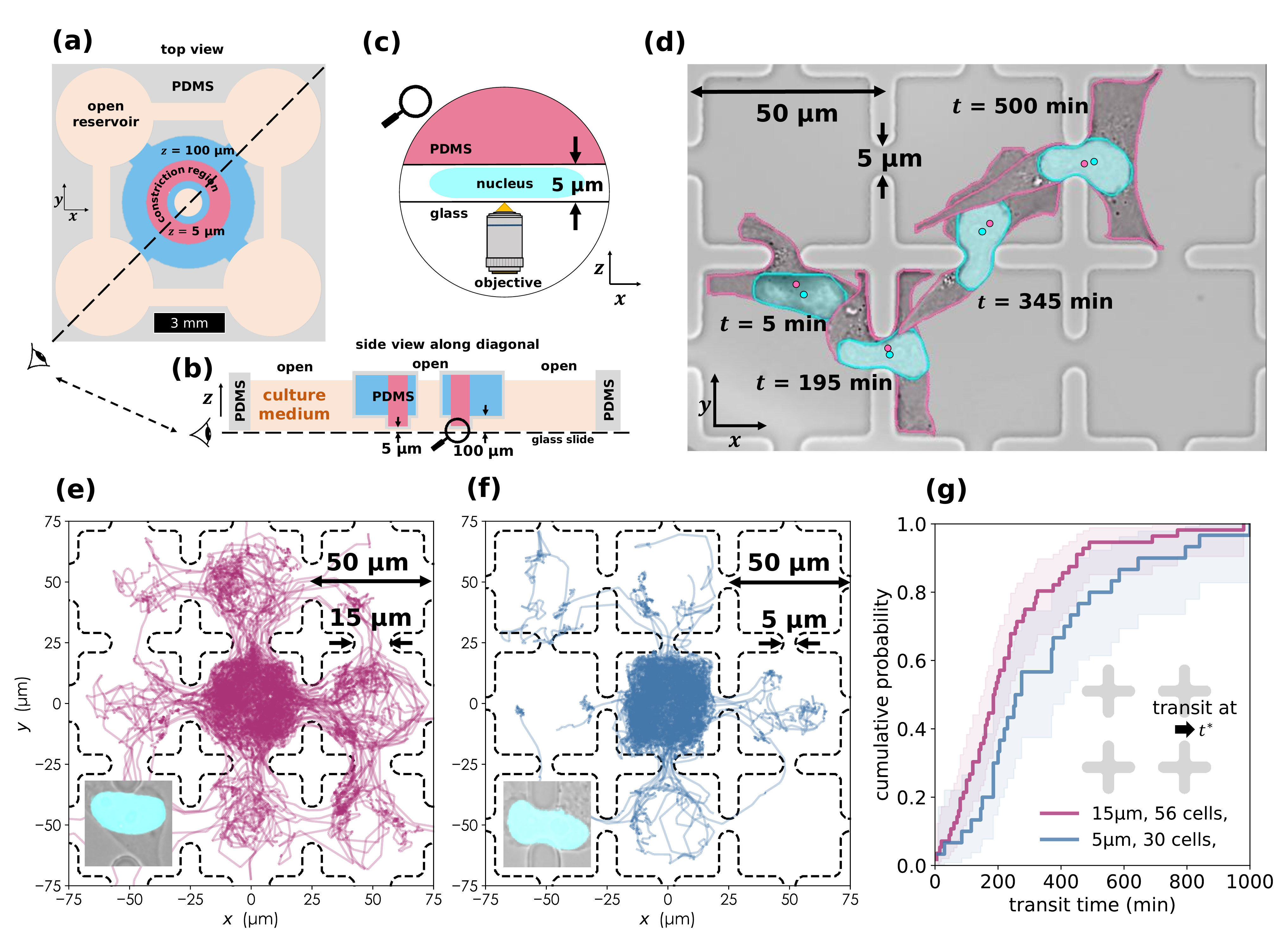}
    
    \caption{\textbf{Experimental setup, trajectories and transit times} \textbf{(a)}Top view scale schematic diagram of PDMS-glass migration device. In the grey region PDMS is adhered to the glass, and the open reservoir in orange is open so that medium can flow in. The region in blue has height $z=100$\um and cells migrate in the pink constriction region where $z=5$\um. \textbf{(b)} side view schematic diagram, not to scale in $z$ axis. \textbf{(c)} Close-up schematic of side view (not to scale). The nucleus is deformed in 5\um space between PDMS and glass in the constriction zone.\textbf{(d)} Example of MDA-MB-231 cell migration in constriction region (brightfield). The cell transits between 50\um$\times$50\um  boxes by passing through 5\um constrictions. Segmented images of the nucleus stained with Hoechst-33342 are overlayed. Cell and nuclear contours are plotted in pink and cyan respectively, along with their centers-of-mass. \textbf{(e)} Trajectories in 15\um constriction pattern (102 cells) and \textbf{(f)} 5\um constriction pattern (131 cells). Brightfield images with nuclear fluorescence overlayad show the nucleus is deformed in 5\um constrictions but not in 15\um. \textbf{(g)} First transit time cumulative probability distributions for cell in 15 and 5\um constriction patterns. The transit time is the time at which the nuclear center of mass has left the initial box. It is only valid if the nuclear center of mass remains outside the initial box for at least 15 minutes.}

    \label{fig:device}
\end{figure*}

We use MDA-MB-231 cells with nuclei stained with Hoechst-33342 dye. Cells migrate spontaneously within the 5\um high PDMS-glass chamber within a microfluidic device. The chamber consists of an array of crosses forming $50\times50$\ums $\,\,$ square boxes separated by a constriction with a width of 15\um or 5\um (\textbf{Fig.~\ref{fig:device}(a-d)}). Cells migrate spontaneously within a box and may transit from one box to a neighboring box through these constrictions. The nuclear radius is on average 7.5\um. As a result, the nucleus of cells migrating through 5\um constrictions is more deformed compared to those migrating through 15\um constrictions (see examples in insets of \textbf{Fig.~\ref{fig:device}(e,f)}). Image acquisition is run over a 22 hour period, with an acquisition rate of one image every 5 minutes. Cell and nuclear contours are obtained, respectively, from brightfield and fluorescence images using a custom trained segmentation model. 

To characterise whether a cell will transit from a box to the neighboring one, we use the position of the nucleus as a proxy. The nuclear position is defined by its center of mass (COM), $\mathbf{r}_{\mathrm{nuc}}(t)=(X_\mathrm{nuc}(t),Y_{\mathrm{nuc}}(t))$ calculated over the surface area defined by the nuclear contour, with equal weighting across the surface. We record the trajectory of the nuclear COM, hereafter referred to as ``trajectory". Trajectories are truncated if any of the following events occur: (1) cell-cell contact; (2) experiment is concluded (after 22 hours); (3) cell leaves the field of view; (4) onset of cell death (5) onset of division.

Given the square symmetry of the boxes, we define the origin of the coordinate scheme as the center of the box occupied by any given cell at the onset of observation. We consider all trajectories as starting in the same initial box (\textbf{Fig.~\ref{fig:device}(e,f)}).
The transit time is defined as the time at which the nuclear COM leaves the initial box, and is only counted as a valid transit if it remains outside the box for at least 15 consecutive minutes. The cumulative probability distributions of transit times are plotted in \textbf{Fig.~\ref{fig:device}(g)}. We observe that for a given time, the transit probability is higher in the 15\um constriction compared to the 5\um ones. Moreover, cells are more successful in migrating through the larger 15\um constrictions, with transit events occurring in 56/102 trajectories as opposed to 30/131 in 5\um.

\subsection*{Defining geometrical features}

The shape-based analysis below requires cell and nuclear contours to be
extracted jointly, which demands segmentation and manual curation of every
frame of every trajectory. We therefore restrict it to two of the four
experiments performed in each condition, chosen independently of transit outcome, and for which both cell and nuclear contours
are extracted. This subset comprises 56 trajectories in the 15\um
condition (31 transits) and 59 in the 5\um condition (8 transits). All analyses from
this point onward refer to this subset of data.

To assess what factors determine whether or not a cell is likely to transit, we define a set of 12 features that reflects the cell and nuclear morphologies, as well as their relative positioning and alignment with the direction of motion. Each feature, obtained from cell and nuclear contours, is calculated for each snapshot in time along a trajectory. A schematic illustration of geometrical notations used for defining features is given (\textbf{Fig.~\ref{fig:distributions}(a)}). 
The 12 features used to describe cell and nuclear morphologies read as follows. To describe cell and nuclear shapes, we use \textit{cell area}, \textit{nuclear area}, \textit{cell perimeter}, \textit{nuclear perimeter}, \textit{cell circularity}, \textit{nuclear circularity}, \textit{cell aspect-ratio}, and \textit{nuclear aspect-ratio}, their mathematical expressions are given \textbf{Tab. \ref{tab:features}}.
To characterize the relative position of nucleus and cell COMs, as well as the orientation of their trajectories, we define \textit{separation}  as the distance between cell COM and nucleus COM, corresponding to the norm of the vector $\mathbf{s}$ (\textbf{Fig.~\ref{fig:distributions}(a)}). \textit{Dot} is the scalar product of the separation vector $\mathbf{s}$ with the nuclear velocity $\mathbf{v}$, \textit{alignment} is the cosine of the angle formed by the separation vector $\mathbf{s}$ and nuclear velocity $\mathbf{v}$. Moreover, \textit{leader projection} is the projection of the velocity on the separation vector $\mathbf{s}$. 

\subsection*{Distributions of feature values}
For each snapshot of a trajectory in the 5 or 15\um constriction condition, the value of each feature is calculated and classified as belonging to a transit or non-transit trajectory. An example of the distribution of values of 4 features on all trajectories for each experimental condition is given \textbf{Fig.~\ref{fig:distributions}(b-e)}. The complete set of feature value distributions can be found in the supplementary information (\textbf{Fig.~S1}). Feature values are widely distributed for all constriction conditions, regardless of whether or not trajectories transit. For most features, distributions for transit and non-transit trajectories are not statistically distinguishable. The only exceptions are \textit{cell circularity} in the 15\um constriction condition, and \textit{cell perimeter} in the 5\um constriction condition, where a significant difference is observed between transit and non-transit cells (\textbf{Fig.~\ref{fig:distributions}(b)} and \textbf{Fig~S1(a)}). In the absence of a clear correlation between most features and transit likelihood, it is possible that useful information is to be found within the complexity of some inter correlation within morphological and dynamic features. To investigate further, we turn to machine-learning approaches. 

\par

\begin{figure*}[tp]
    \centering
        \includegraphics[
            width=\textwidth,
        ]{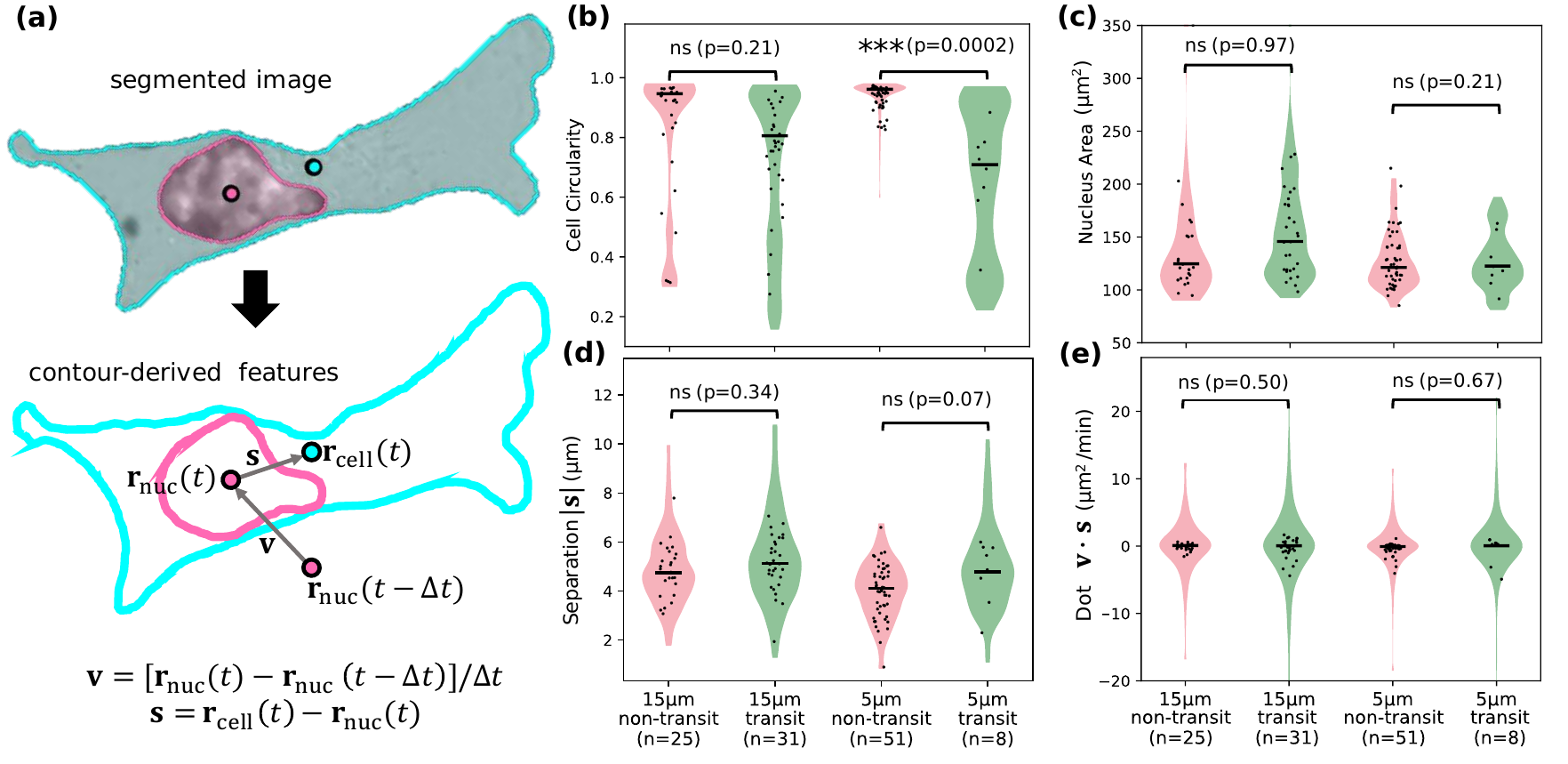}
    
    \caption{\textbf{Contour derived features and examples of their distributions} \textbf{(a)} Schematic diagram of a segmented image used to obtain cell (cyan) and nuclear (magenta) contours with representation of cell and nucleus COM, velocity vector $\mathbf{v}$ and cell-nucleus separation vector $\mathbf{s}$. \textbf{(b-e)} Distributions of contour-derived features for every snapshot of transiting (transit, red) and non-transiting (non-transit, green) cells in 15 and 5\um constriction patterns. \textbf{(b)} \textit{cell circularity}, \textbf{(c)} \textit{nucleus area} \textbf{(d)} \textit{separation} defined as $|\mathbf{s}|$, and \textbf{(e)} \textit{dot} feature defined as  $\mathbf{v}\cdot\mathbf{s}$. The red and green violin area represent data from snapshots at all time frames, whereas the black points represent time-averaged values for individual trajectories. 
    The statistical significance is reported based on the $p$ value from the Mann-Whitney U test performed on time-averaged values to avoid cells tracked for longer times skewing the data.}

    \label{fig:distributions}
\end{figure*}

\begin{figure*}
    \centering
    \includegraphics[width=.9\linewidth]{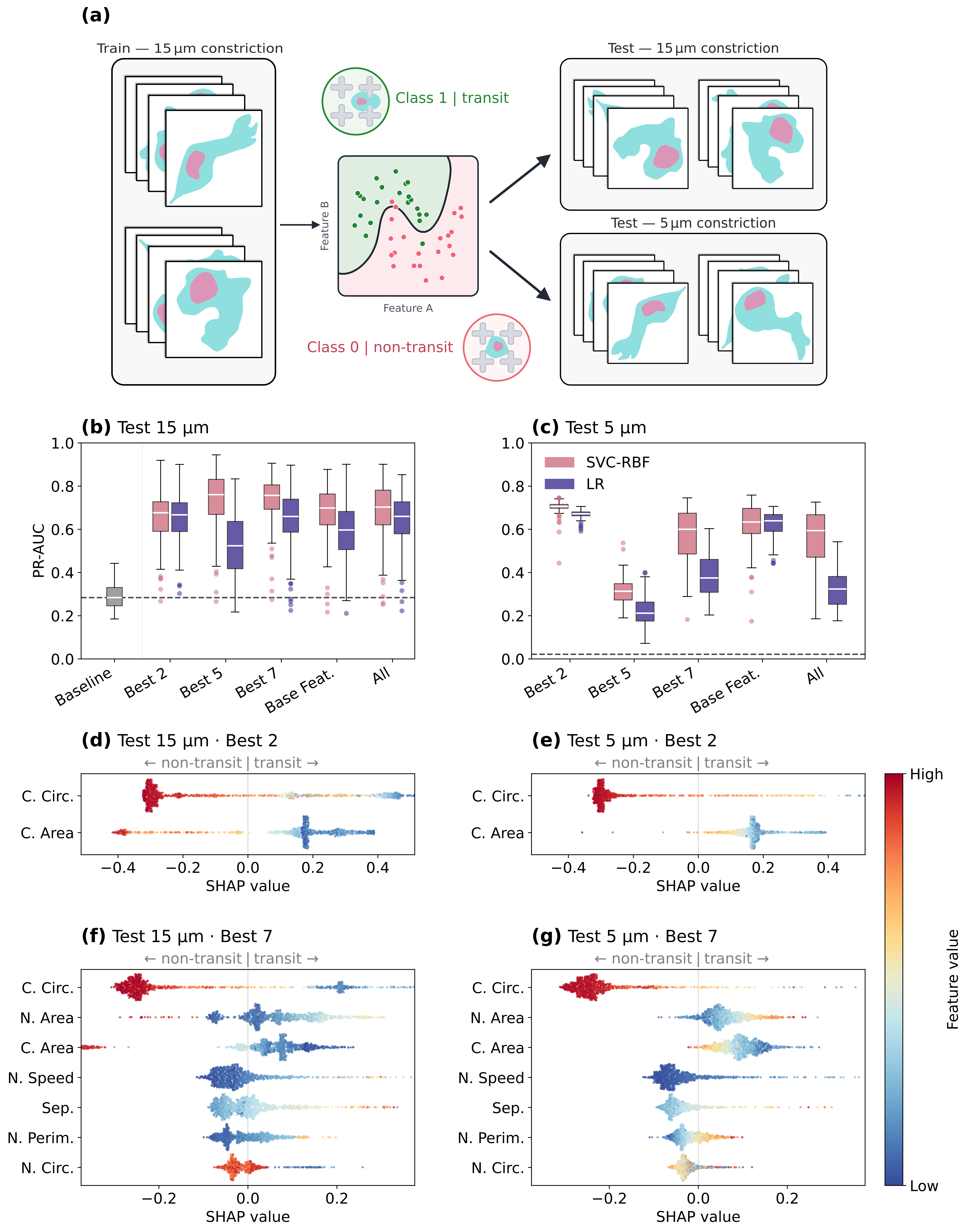}
    \caption{\textbf{Set up of the classification training and results.} \textbf{(a)} 
    Training setup: We first train on trajectories in $15$\um constrictions (cell blue, nucleus pink), and test on $15$\um data, as well as on $5$\um constriction data (bottom panel). \textbf{(b)} PR-AUC box plots for $15$\um test set, for a choice of four different feature sets (\textbf{Tab.~\ref{tab:features}}). Comparison Support Vector Classification with radial basis function kernel (SVC-RBF, pink box) and Logistic Regression (LR, purple box). The baseline (gray box, dashed line) is a result of a random classifier on the $15$\um test set. \textbf{(c)} PR-AUC box plots for a transfer learning on $5$\um test set, for a choice of four different feature sets (\textbf{Tab.~\ref{tab:features}}). Comparison between SVC-RBF (pink box) and LR (purple box) results. The baseline (gray dashed line) is a result of a random classifier on the test set. The spread in the PR-AUC comes from the spread in the results of the model trained on the train test in \textbf{(b)}. Shapley value beeswarm plot for \textit{best 2} feature set for \textbf{(d)} $15$\um  constriction, \textbf{(e)} $5$\um constriction data. Shapley value beeswarm plot for \textit{best 7} feature set for \textbf{(f)} $15$\um  constriction, \textbf{(g)} $5$\um constriction data. The color map depicts the value of the feature with red high and blue low (linear interpolation between $5$th-$95$th percentiles). High cell circularity in both test sets predicts non-transiting trajectories, while low cell area is an indicator of a transiting trajectory.}
    
    \label{fig:classification}
\end{figure*}

\subsection*{Classification task with base and dynamical features}
In addition to the 12 \textit{base features} described above, which are computed for each snapshot acquired at $5$-minute intervals along each trajectory, we compute a set of \textit{dynamical features} as the backward-finite differences of the base features. Furthermore, we also include cell and nucleus velocities.  Altogether, we gather 26 features and use them to predict whether and when trajectories undergo a transit event through a constriction. The trajectories span from the beginning to the end of the measurement and, in some cases, include multiple transitions through successive constrictions. We clip the trajectories into sub-trajectories, defined by a single transition through a constriction. The sub-trajectories then include snapshots spanning from $t_{\min} = 5$ to $t_{\max} = 120$~minutes before the transition, which do not overlap. 
This choice is made by varying the time window while looking at the class balance of the transit and non-transit classes (keeping equal sizes of the two classes in the dataset): choosing a higher $t_{\max}$ includes more transit events, and disturbs the balance of classes. However, we find that different time window choices do not impact the results qualitatively. The set of sub-trajectories for all trajectories in the experiment is what we refer to as a dataset. 

To understand which features determine the probability of a transit through a constriction, we define a binary classification task that predicts from a single snapshot whether a trajectory will undergo a transit event at a later time or not. With all feature values obtained on the $15$\um constriction condition, we train a Support Vector Classifier with a radial basis function kernel (SVC-RBF) to capture potential non-linear relationships in the data, and compare its performance against Logistic Regressions (LR).

\subsection*{Training and feature selection setup} 

The models are evaluated using the area under the precision-recall curve (PR-AUC). Since the $5$\um trajectories contain a small number of transit events ($\sim 14 \%$ of transiting trajectories for our chosen interval and filtering procedure), we train exclusively on the dataset in the $15$\um constriction condition ($\sim 55\%$ of transiting trajectories). The scheme of the set up is shown in \textbf{Fig.~\ref{fig:classification}(a)}. 

To identify the most predictive features, we compare the performance of models trained with different sets of those features. Model selection requires partitioning the dataset into a train set (50\% of trajectories), a validation set (25\%) on which we compare model performance, and finally a test set (25\%) to assess the performance of the best model on held-out data that were not used for either training or model comparison. Both SVC-RBF and LR models additionally have regularization parameters to avoid overfitting (see Methods for how they are set, and \textbf{Tab.~\ref{tab:C-sensitivity}}).

Model performance is assessed using the average PR-AUC over 3 folds of the train and validation sets (75\% together). The space of models to compare corresponds to all possible sets of features, and is thus very large. We use a greedy search to look for the most predictive ones. We first draw a random ordered list of features, and then add features one by one from the top of that list until PR-AUC improves by less than 0.01 upon adding a feature. We repeat the procedure over many random choices of the ordered list. We then group models by the size of their feature sets, and  rank them by their average performance on the validation fold. To avoid fold-specific biases, the top 3 models for each set size are further evaluated by re-computing their PR-AUC over 5 random choices of the split between the 3 folds.

From the greedy procedure emerge three candidate feature sets that we call \textit{best 2}, \textit{best 5} and \textit{best 7} features. The \textit{best 2} feature set is made of  \textit{cell circularity} and \textit{cell area}. These features are also included in the \textit{best 7} feature set, alongside \textit{separation, nuclear speed, nuclear circularity, nuclear area} and \textit{nuclear perimeter}. The \textit{best 5} features are \textit{cell aspect ratio, cell aspect ratio velocity, cell circularity, nuclear area} and \textit{nuclear speed norm}. Note that \textit{nuclear speed} and \textit{cell aspect ratio velocity} are the only dynamical features selected by our procedure for these feature sets.

\subsection*{Testing models obtained using the 15$\mu m$ constriction data} 

The test PR-AUC varies substantially depending on how the data are split. We therefore repeat the evaluation over $100$ random train/test splits and show the resulting distribution of test PR-AUCs as box plots. The spread of each box reflects this split-to-split variability, \textbf{Fig.~\ref{fig:classification}(b)}.  

Among the three feature sets for the $15$\um data, the \textit{best 5}  and \textit{best 7} features have the highest average SVC-RBF PR-AUC with $0.731$ for the \textit{best 5} features, and $0.733$ for the \textit{best 7} features, with  $[0.56, 0.91]$ and $[0.59, 0.88]$ $95\% $ confidence interval (CI) respectively, \textbf{Tab.~S1}. The \textit{best 7} features outperform  the full set of $26$ features (\textit{all}), which include the \textit{dynamical features}, and the \textit{base features}, for both SVC-RBF and LR (see \textbf{Tab.~\ref{tab:feature_sets}} for a summary of the feature sets). The median performance for all models remains $30\%$ above the baseline, see \textbf{Tab.~S1}.

SVC-RBF classifiers compared to LR have a better mean performance, apart from the \textit{best 2} feature set where they match. For all models, we find that the various feature sets contain relevant information about future potential transits through the constrictions, and that, even with a simple linear LR model, the \textit{best 2} feature set already captures relevant signal about the transiting versus non-transiting subsets (\textbf{Fig.~\ref{fig:classification}(b)}). While the greedy search algorithms expanded the feature set to seven features, the top two features, \textit{cell circularity} and \textit{cell area}, capture the primary information about the crossing events. 

\subsection*{Predicting transit events for trajectories in the 5{$\mu m$} constriction}

We find that applying the $15$\um-trained models to the $5$\um condition yields results that outperform the baseline for all models (\textbf{Fig.~\ref{fig:classification}(c)}, random baseline $0.02$), confirming successful generalization of the model. The baseline is very low since there are not many  transit events for this constriction condition compared to the $15$\um dataset. We notice a significant drop in the performance of the \textit{best 5} feature set, compared with the other feature sets, with both its mean SVC-RBF and LR performance below all other feature/model choices (see \textbf{Tab.~S1}). The paired difference in PR-AUC between the models, presented in \textbf{Tab.~S2} shows that the \textit{best 2} features significantly (in terms of CIs) outperforms the \textit{best 5} feature set  (gap in the mean $+0.39$). Overall, for most feature sets, SVC classifiers have better mean PR-AUC than LR suggesting that the nonlinearity of the kernel is important for describing the $5$\um dataset. We conclude that, for the $5$\um constriction data, the \textit{best 2} feature set performs best. 

We note that the PR-AUC evaluated on the test $15$\um data exhibits high variance (visible through the spread of the box plot and the error bars), while the error bars are substantially smaller when evaluated on the $5$\um test data. The box plots in \textbf{Fig.~\ref{fig:classification}(b, c)} differ in spread for two reasons: sampling of the training and test sets, and variation between the trained models themselves. In the $5$\um constriction condition, we separate these effects by keeping the test set constant across the models, trained on different splits. The remaining variation in the PR-AUC in \textbf{Fig.~\ref{fig:classification}(c)} is solely due to differences between the trained models. 

These results lead us to conclude that \textit{cell circularity} and \textit{cell area} generalize well across the two constriction sizes. Additionally, the drop for the \textit{best 5} features suggests that some of its features that are not included in the \textit{best 2} and  \textit{best 7} feature sets, such as \textit{cell aspect ratio}, behave differently in the $5$ \um constriction setup than in the $15$ \um one. 

\subsection*{Evaluating feature importance using Shapley values}
We now know which feature sets predict the transition probability, but not how individual feature values affect the prediction. To assess this, we compute Shapley values using the SHAP package~~\cite{Shapley1953,NIPS2017_7062}, to evaluate the impact of individual features and their values (high or low) on the classification task (prediction of trajectory transit versus non-transit). 
Shapley values evaluate the average effect of a feature on the predicted probability of a transiting trajectory. We  relate each Shapley value to the corresponding feature value (color map: high = red, low = blue)
for the $15$\um and $5$\um constriction setups (\textbf{Fig.~\ref{fig:classification}(d-g)}). 
Each dot represents a single snapshot in the test set and its horizontal position tells about the direction of the feature prediction for that snapshot: right means prediction towards transit, left towards non-transit, and $0$ means no effect.

For the $15$\um constriction setup, \textit{cell circularity and area} are most informative about whether a cell is likely to transit, both for the {\textit{best 2}} (\textbf{Fig.~\ref{fig:classification}(d)}) and {\textit{best 7}}  feature sets (\textbf{Fig.~\ref{fig:classification}(f)}). Round, big cells are less likely to transit than elongated, small cells.  Moreover, very large nuclei, similarly to very large cells, are less likely to transit, while small cells and nuclei are more likely to transit. Yet the relationship between \textit{nuclear area} and transiting is non-monotonic, and medium sized nuclei are also very likely to transit. Low \textit{nuclear speed},  small \textit{nuclear perimeter}, high values of \textit{nuclear circularity} and  low values of \textit{separation} are  weakly associated with non-transiting cells. However large \textit{nuclear speeds} and large \textit{separation} values are strongly associated with transiting cells. Overall, the model predicts  a small elongated cell with a medium sized fast moving nucleus as likely to transit. Round, big cells, with a slow moving nucleus are indicative of non-transiting.  

Both the \textit{best 2} and \textit{best 7} feature sets  mostly rank the influence of features on the cell behavior in the same way for the $5$\um constriction setup as the $15$\um setup, showing that the models use the same logic to discriminate between the classes (\textbf{Fig.~\ref{fig:classification}(e, g)}). The biggest differences occur in the prediction pattern for the \textit{nuclear speed, area} and \textit{nuclear perimeter} (\textbf{Fig.~\ref{fig:classification}(g)}). For the $5$\um data, the \textit{nuclear speed} is always low and \textit{nuclear area} and \textit{perimeter} are monotonic, in contrast to the $15$\um results. For the $5$\um data constriction setup, 
a large nucleus is indicative of transit, whereas for the $15$\um setup medium sized nuclei would indicate transit but big ones would strongly predict non-transit.  In \textbf{Fig.~S2} we show the distribution of feature values, which confirms that while in the $15$\um data (and the $15$\um fibronectin data, discussed below), large \textit{nuclear speed} correlates with trajectories that transit, in the $5$\um data, almost all trajectories that transit have small speed. These results confirm that the nucleus
speed feature does not generalize between the two constriction conditions as we noted when discussing  the PR-AUC in (\textbf{Fig.~\ref{fig:classification}(c)}).

\subsection*{Predicting transit events in fibronectin-coated 15$\mu m$ wide constrictions} 
\begin{figure*}[tp]
    \centering
    \includegraphics[width=\linewidth]{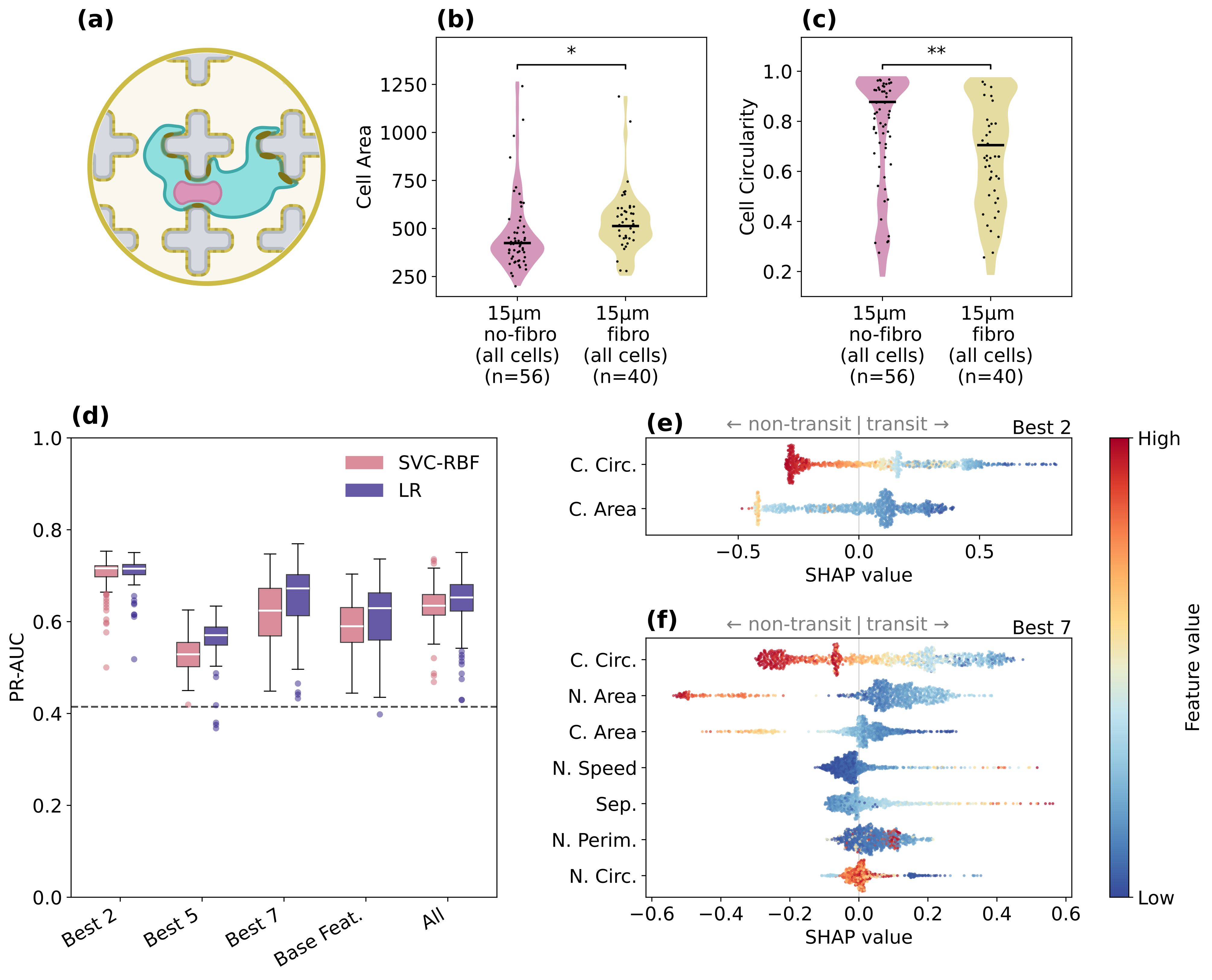}
    \caption{\textbf{Results of the predictions on a test set of trajectories transiting through $15$\um constriction in increased adhesion conditions (fibronectin-coated).} \textbf{(a)} Fibronectin coating leads to higher sticking to the constriction lattice pillars. \textbf{(b)} Violin plot comparison of \textit{cell area} and \textbf{(c)} \textit{cell circularity} between trajectories in non-coated and fibronectin-coated lattice. The statistical significance is reported based on the $p$ value from the Mann-Whitney U test performed on time-averaged values. \textbf{(d)} PR-AUC box plots for a out-of-distribution prediction on $15$\um fibronectin test set, for a choice of five different feature sets (Tab.~\ref{tab:features}). Comparison between Support Vector Classification with radial basis function kernel (SVC-RBF, pink) and Logistic Regression (LR, purple) results. \textbf{(e)} Shapley value beeswarm plot for $15$\um  constriction and fibronectin coating for \textit{best 2} and \textbf{(f)} \textit{best 7} features. The color map depicts the value of the feature with red high and blue low (linear interpolation between $5-95$th percentiles).}
    \label{fig:fibronectin}
\end{figure*}

Fibronectin is a protein that promotes cell adhesion to its surrounding environment, and has long been studied for its role in cancer metastasis and tumorigenesis \cite{lin2019fibronectin,lugo2020breast}. Adherent cells migrate using the mesenchymal mode of migration, resulting in a more elongated morphology \cite{longstreth2024role}. To see whether the models transfer to conditions of enhanced adhesion inducing shape change, we also test our models on trajectories in a fibronectin-coated microfluidic device in 15\um constriction condition. Cells in fibronectin-coated constrictions have larger cell area and lower circularity compared to the non-coated substrates \textbf{Fig.~\ref{fig:fibronectin}(b,c)}. 

All models perform above the baseline ($0.42$), with the \textit{best 2} feature set having the highest SVC-RBF/LR mean PR-AUC  $0.70$/$0.71$, with $[0.45, 0.90]$/$[0.47, 0.90]$ $95\%$ CI~(\textbf{Fig.~\ref{fig:fibronectin}(d)}). 
Although the confidence intervals of the larger feature sets overlap with the baseline, their means are greater than $0.53$, confirming that the models trained on non-treated substrates generalize to the fibronectin data. Interestingly, for all feature sets, LR works better than SVC, suggesting that the non-linear kernel does not help separate the classes and may instead overfit. Here again, analogously to the results for the $5$\um constriction data, we notice that the \textit{best 5} feature set performs on average worse than other feature sets. This observation suggests that perhaps individual features, in particular the \textit{cell aspect ratio}, do not generalize well to different constriction and adhesion settings. 

The analysis of the Shapley values for the \textit{best 2} and \textit{best 7}  feature sets show similar results to those observed in the non-coated experiments and the $5$\um data, (\textbf{Fig.~\ref{fig:fibronectin}(e, f)}). Overall, the model sees an elongated cell with a small nucleus as a transiting cell, in particular if the \textit{separation} and \textit{nuclear speed} are high (\textbf{Fig.~\ref{fig:fibronectin}(f)}). We observe the same behavior of the \textit{nuclear speed} as for the $15$\um trained model. 
While the increased adhesion has a significant impact on the individual values of the features, which include the cell and nucleus geometry and dynamics, (\textbf{Fig.~\ref{fig:shap-feature-comparison-15-5-fibro}}), the $15$\um trained model retains its predictive power and generalizes successfully to the increased adhesion data. These results suggest that the transit probability is a result of a complex and non-linear dependence on cell and nucleus geometry and dynamics.

\subsection*{Prediction of transit times} 

After identifying features that separate trajectories that transit from the ones that do not, we now evaluate whether we can predict the time at which a trajectory will transit through a constriction. In the following, we start using the \textit{best 7} feature set, while results for all other feature sets are presented in \textbf{Fig.~S3}.

We reformulate the time-to-transit task into a family of binary classification tasks parametrized by a split point $t_c$: for a snapshot recorded at time $t$ in a trajectory that transits at time $t^* \geq t$,  $\Delta t = t^* - t$ is the time remaining until transit. The snapshot is labeled  \textit{near-transit}, if $0 \leq \Delta t \le t_c$ and  \textit{distant} if $\Delta t > t_c$. We ask at what value of $t_c$ the features of the trajectories are most predictive about the transit times. We vary the split point $t_c$ between $[5, 120]$~minutes (in $5$~minutes steps) and train the SVC-RBF and LR classifiers (see Methods) using the set of features from the previous binary classification analysis.

We train the model on data from $15$\um constriction, restricting the trajectories to the transiting ones only, with some trajectories having more than one transit event. To ensure that models are always tested on held-out data, we train them on 3 folds of the transiting trajectories and test on the remaining one (corresponding to the $75\%/25\%$ train/test split). This procedure is repeated over $100$ random choices of the split between folds.
Furthermore, to compensate for the class imbalance between the \textit{near-transit} and \textit{distant} classes for different choice of the time split $t_c$, each training observation of class $a$ was weighted by $w_a = N_\mathrm{train}/(2N_{\mathrm{train},a})$, where $a =\textit{near-transit}$ or ${distant}$. Feature values were standardized to zero mean and unit variance using statistics computed on the training data only. 

Discriminability between the \textit{near-transit} and \textit{distant} classes is quantified by the mutual information $I(y;\hat y)$ between true and predicted labels, where the labels are the classes assigned to each snapshot, \textbf{Fig.~\ref{fig:ml-time}(a)}. The mutual information is bounded by the entropy of the true labels, $H(y)$, which reflects how hard it is to guess the label without any information. $H(y)$ is zero if all data belong to one class, and largest when the two classes are equally common. For the two classes (\textit{near-transit} and {distant}), $H(y) = -p\log_2 p - (1-p)\log_2(1-p)$, where $p$ is the fraction of \textit{near-transit} trajectories, and it varies with $t_c$. $H(y)$ reaches a maximum of $1$~bit for $p = 0.5$ at $t_c=45$~minutes. For this $t_c$ the two classes are perfectly balanced (see \textbf{Fig.~S4}). For details on the computation of the mutual information, see Methods. 

To separate the classification performance from the effect of the class balance, we report the normalized mutual information $U=I(y;\hat y)/H(y)$, \textbf{Fig.~\ref{fig:ml-time}(b)}. As the baseline we use a null model in which, for each $t_c$, we randomly shuffle the transit times in the training set before labeling them into \textit{near-transit/distant}. The model is then retrained on these shuffled labels and its predictions are scored against the true labels of the test cells. We train the model for the SVC-RBF (blue), LR (red) and the null (dashed line, blue and red for both classifiers respectively). Solid lines in \textbf{Fig.~\ref{fig:ml-time}(a,b)} show the mean over the $100$ cross validation repetitions and shaded bands depict the standard deviation for the \textit{best 7} feature set.  

For both classifiers, SVC-RBF and LR, the mutual information and its normalized version are non-monotonic in $t_c$, with a maximum at $t_c=30$~minutes. The two classifiers yield nearly identical curves and are consistently above the null model. The classifiers  identify a persistent pre-transition signal over the whole range of $t_c$, which is most pronounced around $30$~minutes. Since the normalized mutual information corrects for class balance, and the peak at $30$~minutes differs from the point of maximal balance $t_c \approx 45$~minutes, we conclude that the maximum is not simply an artifact of the high label entropy  when classes are balanced.

For the other feature sets, the normalized mutual information is also highest around $t_c=30$-$35$~minutes (\textbf{Fig.~S3}). These results show that cell and nucleus features carry information about the time to transit over most of the two hours preceding the transit event, most likely with the clearest signature about half an hour before transit.

\subsection*{Testing the model on increased adhesion conditions and narrower constriction}
\begin{figure*}[tp]
    \centering
   \includegraphics[width=0.7\linewidth]{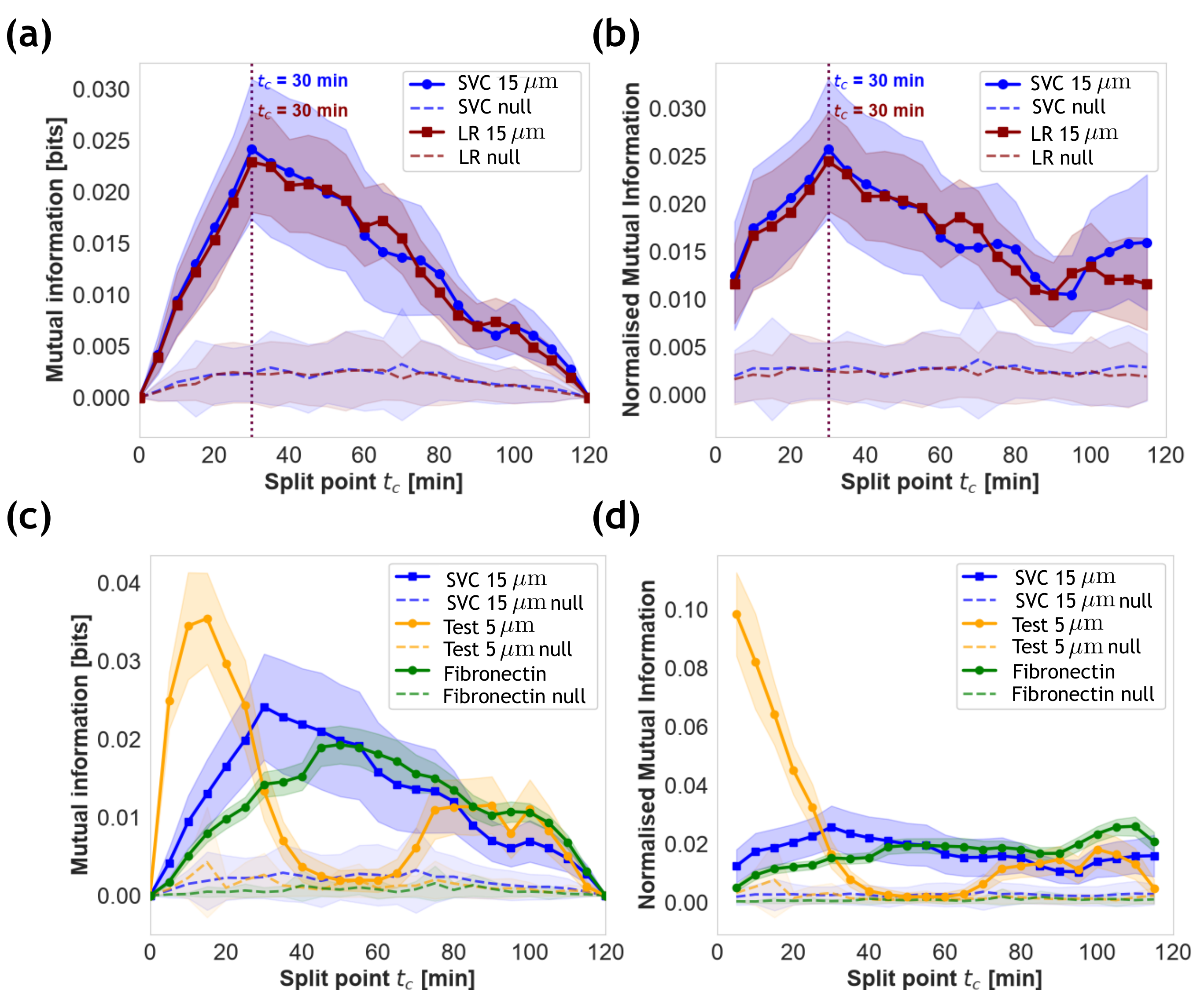}
    \caption{\textbf{Mutual information for the prediction of transit times.}
    The classifiers Support Vector Classifier with radial basis function kernel (SVC, blue) and Logistic Regression (LR, red) are trained $15$\um constrictions using the \textit{best 7} feature set. \textbf{(a)} Mutual information $I$ between true and predicted class as a function of the split point $t_c$ for the classifiers tested on the $15$\um constrictions trajectories. Solid lines show the mean over the $100$ repetitions and shaded areas the standard deviation. Dashed lines show the null model, which is the same classifier trained on $15$\um data with randomly shuffled transit time labels. \textbf{(b)} Normalized mutual information. In \textbf{(a)} and \textbf{(b)}, the mean of both classifiers is highest at $t_c=30$~minutes (dotted line). \textbf{(d)} Mutual information, \textbf{(e)} normalized mutual information for the SVC-RBF \textit{best 7} trained on $15$\um constrictions and tested without retraining to $5$\um constrictions (yellow) and to $15$\um constrictions coated with fibronectin trajectories (green). For comparison in blue is the the $15$\um result from \textbf{(a-b)}. Dashed lines show the corresponding null models (for $5$\um and fibronectin: mean and standard deviation over $20$ repetitions).
}
    \label{fig:ml-time}
\end{figure*}

Analogously to the previous analysis, we test our SVC-RBF model from \textbf{Fig.~\ref{fig:ml-time}(a)} (\textit{best 7}) on transiting trajectories in $5$\um and $15$\um fibronectin-coated constrictions, to verify whether the learned morphology signature generalizes beyond the training geometry. We apply the classifier without any retraining to trajectories in $5$\um and on $15$ \um fibronectin-coated. All trajectories of both conditions are used in the test set, and no cross validation is required. 
Features of each test condition are standardized with its own mean and standard deviation. 

The results of the SVC-RBF classifier (\textit{best 7}) tested on the $5$\um and $15$\um fibronectin-coated constrictions are presented in \textbf{Fig.~\ref{fig:ml-time}(c-d)} (yellow and green line respectively, blue line SVC-RBF on $15$\um for reference). Analogously to the analysis on the $15$\um trajectories, the baseline is the null model obtained from the $15$\um training set evaluated on the $5$\um and $15$\um fibronectin test set (yellow and green dashed line respectively). In the $5$\um constriction condition, the mutual information fluctuates and drops to the null level for $t_c = 40-70$~minutes, indicating that the predictions carry no information about the transit events in this range. In contrast, the SVC-RBF classifier trained on $15$\um constriction trajectories generalizes well when tested on all available data for $15$\um fibronectin-coated constrictions. The normalized mutual information is above the null level for all $t_c$ and follows a trend similar to that of the non-coated $15$\um data, with the similar average ($0.016$ vs $0.017$). The normalized mutual information has no single clear significant peak but is mostly elevated for $t_c > 30$~minutes. We conclude that the \textit{best 7} feature set carries the most information about the time to transit for trajectories in the fibronectin-coated constrictions more than half an hour before transit. Together, these results show that the pre-transition signal learned on $15$\um constriction trajectories carries over to $15$\um fibronectin-coated data but not to the $5$\um setup. 

\section*{Discussion}

We present a microfluidic device in which 2D-projected trajectories are measured for cells that experience transit events as their nuclei pass through constrictions. The advantage of this geometry is that cell and nuclear contours can be measured in regions of relatively low confinement before cells enter constrictions. Of the base set of features we define, only \textit{cell circularity} and \textit{cell perimeter} show statistically distinguishable distributions for trajectories that transit or do not (see distributions of feature values in \textbf{Fig.~\ref{fig:distributions}} and \textbf{Fig.~S1}). 

To extend the analysis beyond the comparison of single feature distributions, we introduce a machine-learning approach that combines many cell and nuclear features. The machine-learning classifiers show that features which distributions do not differ significantly between transiting and non-transiting cells on their own, such as \textit{cell area}, \textit{separation} and \textit{nuclear speed}, do contribute to the prediction when combined, and that transit can be predicted well above chance. Furthermore, Shapley values rank the features by their importance and reveal how their values, and non-monotonic effects, can change the probability of transit. For $15$\um constriction data, the model sees an elongated cell with a large, fast moving nucleus as a sign of transit. In contrast, round, big cells, with a slow moving nucleus are indicative of non-transiting. 

Round morphology is characteristic of the amoeboid mode of migration, whereas elongation is characteristic of mesenchymal migration, where cells are more strongly adhered to the substrate. While it has been reported that amoeboid migration can lead to higher migration speeds in confined environments~\cite{holle2019cancer,gabbireddy2021melanoma}, our results indicate that this strategy could be less efficient in going through constrictions in the geometry we study. 

The classification task of whether trajectories will transit through $15$\um constrictions generalizes well to two other distinct experimental conditions, such as smaller constrictions, and surfaces coated with adhesive fibronectin. Note that \textit{cell circularity} is the only feature which distribution is most significantly different between transit and non-transit classes in $15$\um constriction condition,
whereas the model is able to find additional ones predictive of transit of non-transit. Similarly, in the $5$\um  condition, more features are found predictive by our analysis, and which are only partially geometrically linked to \textit{cell perimeter}, the only most significant feature value that experimentally distinguishes transit from non-transit cells. Although the coating significantly changes the distribution of feature values compared to non-coated conditions (see \textbf{Fig.~S3}), the model retains its predictive power with \textit{cell circularity} and \textit{cell area} being the most important, in agreement with the value distributions being significantly different. All together, we find that the model trained on $15$\um constriction condition has the power of detecting additional features that carry information about the probability of transit in the two other conditions. The model transfers well to $5$\um data, with some differences in the Shapley values compared to the $15$\um constriction data. In the $5$\um data, \textit{nuclear speed} is almost always low, so that the association of a fast nucleus with transit seen at $15$\um does not occur, while \textit{nuclear area} and \textit{nuclear perimeter} act monotonically, with larger values favoring transit. This may reflect the effect of the active cytoskeleton that has more surface to hold on the nucleus to drive it through smaller constrictions. Because the velocity is at its highest in the constriction~\cite{amiri2025nuclear}, only a very few snapshots of the data sets correspond to the nucleus in constrictions. Therefore, larger values of nuclear size are not an artefact of the experimental setup geometry, but rather a real feature of cell capacity to migrate through constrictions. 

For the prediction of transit times, we reformulate the task as a binary classification of whether a trajectory will transit within a given time interval. Both SVC-RBF and LR distinguish \textit{near-transit} from \textit{distant} time points above chance throughout the two hours preceding transit, indicating a persistent pre-transit signature in cell and nucleus morphology and dynamics. The information about the transit time is most likely highest for $t_c$ around half an hour before transit. The SVC-RBF classifier trained on $15$\um constriction conditions with the \textit{best 7} features set generalizes to fibronectin-coated $15$\um data, where it performs above chance at nearly all $t_c$. Unlike in the uncoated conditions, the normalized mutual information shows no clear maximum around $30$~minutes but is elevated for all $t_c$ beyond about $30$~minutes before transit. This suggests that, in the fibronectin-coated setup, features show signatures of a transit in a period longer than $30$~minutes before transit. In contrast, on the narrower $5$\um constrictions, the normalized mutual information drops to the null level for $t_c = 40$-$70$~minutes, meaning that the predictions are uninformative in this range. Perhaps due to the deformation that cells have to undergo in narrower constrictions, features learned to be predictive on $15$\um dataset do not mark the approaching transit in the same way.

In past work, MDA-MB-231 cells in extra-cellular matrix were classified into morphological phenotypes based on geometrical features extracted from cell contours~\cite{eddy2021morphodynamics}. The authors quantify the ``dwell-time" as the time cells spend in one morphological state before transforming into another. For some morphological phenotypes, this dwell-time is approximately $30$-$60$~minutes. This gives a timescale at which cells change their morphologies. The most likely position of the maximum we observe, $t_c\approx30$~minutes, is consistent with this time scale. After about $30$~minutes, a cell may have changed its morphology, so that its features at time $t$ become less predictive of a transit at later times. 

In summary, our results indicates that the probability of transit depends on cell and nucleus geometry and dynamics in a complex, non-linear way. Considering sets of features rather than single features captures this dependence and allows transit to be predicted successfully..

Beyond the geometric features used here, our framework can be applied as a
systematic way to identify which cell properties determine its ability to migrate through confined spaces. Any quantity that can be measured in
time in live cells can be added as an input feature, and the classifier then
ranks it by how well it predicts transit. This is useful because many
candidate determinants of migratory capacity are currently studied
individually, so their relative importance for confined migration is
unknown. Examples include chromatin compaction~\cite{hsia2022confined}, the
amount of nuclear scaffolding proteins~\cite{tran2025cytoplasmic},
condensates~\cite{todorovski2023confined}, the positioning of the
Golgi apparatus~\cite{vaidvziulyte2022persistent}, and the activity of signaling
proteins such as Rac1 and Cdc42, which influence cell morphology and
migration~\cite{yamao2015distinct}. Imaging several of these simultaneously
and including them as features would allow a direct, quantitative comparison
of their predictive power.

\section*{Methods}
\paragraph*{Cell culture.}
\label{app:culture}

MDA-MB-231 cells are cultured in DMEM (Dulbecco’s Modified Eagle Medium – Gibco) with 10\% volume FBS (Fetal Bovine Serum – Gibco) at 37$^{o}$C in a humidified incubator with 5\% CO$_2$. Cells were passaged twice a week with a 1:4 dilution for each passage. Cells were not used beyond 14 passages.

\paragraph*{Microfluidic device assembly.}
\label{app:assembly}
 
Polydimethylsiloxane (PDMS) and crosslinker (Dow Sylgard 184 Elastomer Kit) are mixed at a
ratio of 10:1 and degassed for 30 minutes in a vacuum desiccator to remove bubbles. The mixture is poured directly onto the silicium wafer (fabricated using photolithography at Institut Pierre-Gilles de Gennes by Taha Messlmani) and placed in a 60$^o$C oven for 6 hours to cure. Individual patterns are cut from the PDMS with a scalpel and the openings are cut with a 3mm and 0.5mm biopsy punches for the medium reservoirs and central aspiration port respectively. Glass coverslips are left to soak overnight in 0.1M HCl solution, then rinsed with deionised H$_2$O, isopropanol, and dried with Kim wipes. The PDMS pattern and coverslip are treated in a plasma cleaner for 1.5 minutes before sticking the pattern onto the glass. The assembled device is placed on a hotplate at 100$^o$C for 20 minutes to finalize the sticking process. Devices are used for migration experiments not more than 2 days after their assembly.

\paragraph*{Cell migration experiment.}
\label{app:experiment}

Devices are sterilized by rinsing twice with a mixture of 70\% EtOH and 30\% H$_2$O (by volume), twice with Phosphate-buddered saline (PBS - Gibco), and twice with cell culture medium (DMEM - Gibco). The device is filled with a suspension of cells in medium at a concentration of $10^6$/mL. Using a 25$\upmu$L Hamilton syringe, 5$\upmu$L volume is gently aspirated through the central aspiration port, drawing cells into the constriction pattern area (see \textbf{Fig.\ref{fig:device}a}). The medium is removed and replenished with medium containing 1$\upmu$g/mL Hoechst 33342, and incubated for 20 minutes before finally rinsing twice with regular medium. After 5 hours incubation, the medium is replaced with DMEM without phenol red and with 15 mM HEPES (Gibco), 100 units/mL penicillin, and 100 $\upmu$g/m: streptomycin (Life Technologies). For experiments with fibronectin, coating is performed on the previous day. For fibronectin coating, assembled devices are treated for 5 minutes in the plasma cleaner, then rinsed twice in 70\% EtOH, 30\% H$_2$O, and once with PBS. Each device is filled with 100$\upmu$L fibronectin in PBS (100 $\upmu$g/mL) and incubated at 37$^o$C for 2 hours in a petri dish sealed with parafilm. The dish is transferred to a 4$^o$C until the following morning. The device is then loaded in the usual way, skipping two EtOH rinsing and the first PBS rinsing steps.

\paragraph*{Microscopy.}
\label{app:microscopy}

All timelapse acquisitions were performed using a Nikon Ti-E epifluorescence microscope fitted with a sCMOS camera (2048 ORCA Flash 4.0 V2, Hamamatsu or Prime BSI, Teledyne), 60x Nikon oil objective, Nikon perfect focus system, and environmental controller set to 37$^o$C, 5\% CO$_2$, with humidification. Imageing is performed using the brightfield and epifluorescence at 360nm LED excitation for the nucleus. Each frame is imaged every 5 minutes, over a period of 22 hours. 

\paragraph*{Image processing.}
\label{app:image processing}

Images are treated using a series of custom Python scripts. After performing a drift correction, cell and nucleus contours are detected using a YOLO-SAM model~\cite{yolov8} adapted to our data by re-training on manually annotated experimental contours. All coordinates are shifted such that the origin is placed at the center of the box the cell occupies initially. The boundary pixel values are used to define cell and nuclear contours. Contours are further processed by spline fitting and downsampling the contour such that the spacing between points is always 1\um. 

\paragraph*{Feature definition.}
The set of 12 features is defined in \textbf{Tab.~\ref{tab:features}}. Features 1 through 8 are based directly on the cell and nucleus contours, and the formula is identical for cell and nucleus. Thus the calculation is presented as for any contour with $n$ boundary points $\omega_j(j=1,...,n)$. Enclosed within the contour are $N$ pixels with coordinates $(x_i,y_i)$. The remaining features (9-12) are derived from the separation vector $\mathbf{v}=\mathbf{r}_{\mathrm{cell}}-\mathbf{r}_{\mathrm{nuc}}$ and the velocity vector $[\mathbf{r}_{\mathrm{cell}}(t)-\mathbf{r}_{\mathrm{nuc}}(t-\Delta t)]/\Delta t$.

\begin{table*}[!t]
\centering
\caption{\textbf{Definition of contour-derived features}. For features 1--8, the same calculation is performed for the cell and nucleus contours, so one formula is given for area, perimeter, circularity and aspect ratio. Formulas apply to any contour with $n$ boundary points $\omega_j(j=1,...,n)$. The boundary points are ordered and closed, such that $\omega_1=\omega_{n+1}$. Enclosed within this boundary are $N$ pixels $\Omega_i$, and each pixel has coordinate vector $\mathbf{r}_i=(x_i,y_i)$. The center of mass has coordinates $(X,Y)$ and is calculated as $(1/N)\Sigma_{i}^{N}\mathbf{r}_i$. $\mathbf{r}_{\mathrm{cell}}=(X_{\mathrm{cell}},Y_{\mathrm{cell}})$ and $\mathbf{r}_{\mathrm{nuc}}=(X_{\mathrm{nuc}},Y_{\mathrm{nuc}})$ are the centers of mass of the cell and nucleus. Features 9--12 are derived from the vectors $\mathbf{s}$ and $\mathbf{v}$ (see \textbf{Fig.\ref{fig:distributions}a}). The vector $\mathbf{s}$ is the difference between cell and nucleus COM, $\mathbf{r}_{\mathrm{cell}}-\mathbf{r}_{\mathrm{nuc}}$. The velocity vector $\mathbf{v}$ is the velocity of the nucleus $[\mathbf{r}_{\mathrm{cell}}(t)-\mathbf{r}_{\mathrm{nuc}}(t-\Delta t)]/\Delta t$.} \vspace{0.3cm}
\begin{tabular}{c l c l}
\hline
\# & feature & units & formula \\
\hline
1  & Cell area (C.\ Area) & \ums & \multirow{2}{*}{$N\times\mathrm{(pixel\,\,area=0.047}$\ums)}\ \\
2  & Nuclear area (N.\ Area) & \ums & \\
\hline
3  & Cell perimeter (C.\ Perim.) & \um & \multirow{2}{*}{$\sum_{1}^{n} |\omega_{j+1}-\omega_j|$} \\
4  & Nuclear perimeter (N.\ Perim.) & \um & \\
\hline
5  & Cell circularity (C.\ Circ.) & -- & \multirow{2}{*}{$4\pi\,(\text{Area})/(\text{Perim.})^{2}$} \\
6  & Nuclear circularity (N.\ Circ.) & -- & \\
\hline
7  & Cell aspect ratio (C.\ Asp.) & -- &
\multirow{2}{*}{\begin{tabular}[c]{@{}l@{}}$R_x/R_y$, where \\ $R_x=\sqrt{\sum_{i=1}^{N}(x_i-X)^2/N}$, $R_y=\sqrt{\sum_{i=1}^{N}(y_i-Y)^2/N}$\end{tabular}} \\
\vspace{0.4cm}
8  & Nuclear aspect ratio (N.\ Asp.) & -- &\\
\hline
9  & Separation (Sep.) & \um & $|\mathbf{s}|=|\mathbf{r}_{\mathrm{cell}}-\mathbf{r}_{\mathrm{nuc}}|$ \\
10 & Dot & \ums/min & $\mathbf{s}\cdot\mathbf{v}$ \\
11 & Alignment (Align.) & -- & $\mathbf{s}\cdot\mathbf{v}/(|\mathbf{s}||\mathbf{v}|)$ \\
12 & Leader projection (L.\ Proj.) & \um & $\mathbf{s}\cdot\mathbf{v}/|\mathbf{v}|$ \\
\hline
\end{tabular}
\label{tab:features}
\end{table*}

\begin{table*}[tp]
\centering
\caption{Feature sets used by the classifiers and their corresponding features and the regularization parameter $C$ found in the feature selection procedure. A small $C$ gives a simpler, more strongly regularized model that tolerates some training errors, while a large $C$ fits the training data more closely but risks overfitting.}
\begin{tabular}{lll}
\hline
\textbf{Feature set} & \textbf{Features} & \textbf{$C$ (LR/SVC)} \\
\hline
Best 2    & Cell area, cell circularity & $10/0.01$\\ 
Best 5 &  Cell aspect ratio, cell aspect ratio velocity, cell circularity, \\&nuclear area,  nuclear speed norm & $0.001/10$\\
Best 7    & Cell area, cell circularity, nuclear area, nuclear circularity,\\           
          & nuclear perimeter, nuclear speed norm, separation & $0.01/1$\\
Base features & \textit{Best 7} without nuclear speed norm + alignment, cell aspect ratio, cell perimeter, dot,\\
          & leader projection, nuclear aspect ratio & $ 0.01/1$\\
Dynamical features      
& cell area velocity, nuclear area velocity, cell speed norm, \\
& nuclear speed norm, cell perimeter velocity, nuclear perimeter velocity, \\
& cell circularity velocity, nuclear circularity velocity, separation velocity,

\\
& dot velocity, cell aspect ratio velocity, nuclear aspect ratio velocity,  \\
& alignment velocity, leader projection velocity &\\
All & Base + Dynamical features & $0.01/1$ \\
\hline
\end{tabular}
\label{tab:feature_sets}
\end{table*} 

\paragraph*{Model inference.}
Model training and inference were conducted using the scikit-learn package~\cite{scikit-learn}. We compare the results obtained with two different models: a Support Vector Classifier with a radial basis function kernel $K(\mathbf{x},\mathbf{x}')=\exp(-\gamma\|\mathbf{x}-\mathbf{x}'\|^2)$ (SVC-RBF), and logistic regression (LR). The kernel parameter $\gamma$ was chosen by the median heuristic, $\gamma = 1/\left(2\,\mathrm{median}_{i<j}\|\mathbf{z}_i-\mathbf{z}_j\|^2\right)$, where $\mathbf{z}_i$ are the standardized feature vectors and the median runs over all pairs of observations in the data set. For both classifiers, the regularization parameter $C$ was tuned by cross-validation on the PR-AUC, computed with \texttt{sklearn.metrics.average\_precision\_score}.

\paragraph*{Feature selection.}

We construct feature sets by greedily looping over permutations of the features. First step: generate a permutation of the feature list. Train the model given on the first feature. Then train the model on the first and second feature. If the score increase is greater than $\varepsilon= 0.01$, iterate, else, terminate and stop the loop here. Here the score is the average, over three-fold cross validation, repeated over $5$ splits, held-out PR-AUC ($75-25\%$ train-test split). Sets were evaluated through numerous splits/repeats.

At each step of the feature selection, models are trained and evaluated for $C \in \{ 0.01, 0.1 , 1\}$. Each training set is associated to its best-performing $C$, defined as the best average on the $3$-fold cross-validation set.

In Methods we describe how we choose the kernel parameter $\gamma$ and the regularization parameter $C$, which controls how closely the model fits the training data (see \textbf{Tab.~\ref{tab:C-sensitivity}}). A small $C$ gives a simpler, more strongly regularized model that tolerates some training errors, while a large $C$ fits the training data more closely but risks overfitting.

\paragraph*{Interpretability.}
Feature importance is estimated using the \emph{Shapley value} ~\cite{Shapley1953}. Let $U$ be the universe of players, the \emph{value function} $v : U \mapsto \mathbb{R}_+$ assigns to $S \in U$ the sum of the total payoff $v(S)$ the coalition of players $S$ can obtain by collaborating. The Shapley value $\phi_i$ assigns to each player $i \in N$ its contribution to the payoff averaged over all possible coalitions $i$ is participating in:

\begin{align*}
\phi_i(v)
&=
\sum_{S \subseteq N \setminus \{i\}}
\frac{|S|!\,(n-|S|-1)!}{n!}
\left(v(S \cup \{i\}) - v(S)\right) \\
&=
\frac{1}{n}
\sum_{S \subseteq N \setminus \{i\}}
\binom{n-1}{|S|}^{-1}
\left(v(S \cup \{i\}) - v(S)\right).
\end{align*}

We compute Shapley values using \texttt{shap.KernelExplainer} from the SHAP package ~\cite{NIPS2017_7062}.

\paragraph*{Filtering time points.}
We drop the last $t_{\max}$ minutes of every trajectory, as these sub-trajectories could be falsely associated to a non-observed transition. Furthermore, for transiting cells, we drop time points to insure every pre-transit time window only includes a single transit event.

\paragraph*{Transit-time classification.}
\label{methods-regression}
We analyze $N_\mathrm{cells}=29$ trajectories on $15\,\mu$m constriction that underwent at least one transition. For each transition we extract all observations (frames) recorded $T\in\{5,10,\dots,120\}$~minutes before it, requiring that no other transition of the same cell occurs in between. In total we have $N=1446$ snapshots. Depending on the choice of the feature set, each snapshot $i$ is described by a feature vector $\mathbf{x}_i$ of its morphological and alignment features (\textit{best 7} in the main text), its time to transition $T_i$, and the identity $c(i)$ of the trajectory it belongs to.\par
For the task of predicting \textit{near-transit} and \textit{distant} snapshots we scan bins marked by $t_c=5,10,\dots,115$~minutes. The mutual information was set to zero at $t_c=0$ and $t_c=120$~minutes, where only one class (\textit{near/distant}) exists.
We use two classifiers, both as implemented in scikit-learn: a Support Vector Classifier with a radial basis function kernel $K(\mathbf{x},\mathbf{x}')=\exp(-\gamma\|\mathbf{x}-\mathbf{x}'\|^2)$ (SVC-RBF), and an $L_2$-regularized logistic regression (LR). For SVC-RBF, $\gamma$ is set by the median heuristic as described in \textit{Model inference}, computed once per feature set on the $15$\um data set. It uses no information about the transit times.\par

In contrast to the transit/no-transit classification, where $C$ was tuned by cross-validation on PR-AUC, here we used a fixed regularization parameter $C=1$ for both classifiers. The transit-time classification task differs in data, labels and class balance (only observations of transiting cells, labeled \textit{near-transit} or \textit{distant} at a varying split point $t_c$), so the values tuned for the transit task do not carry over. Table~\ref{tab:C-sensitivity} shows that the results are insensitive to $C$ over a broad range: for SVC-RBF, mutual information and PR-AUC varied only moderately for $C \in [0.1, 10]$ for all feature sets, whereas for $C \le 0.01$ the mutual information dropped to nearly zero, i.e.\ the predicted labels carried almost no information about the true labels. LR was insensitive to $C$ over $[10^{-3},10^{2}]$. The choice $C=1$ thus lies well within the range of stable performance for all feature sets and both classifiers.\par

To compensate for the class imbalance, which varies strongly with $t_c$, each training observation of class $a = \textit{near-transit/distant}$ was weighted by $w_a = N_\mathrm{train}/(2N_{\mathrm{train},a})$. Features are standardized to zero mean and unit variance using statistics computed on the training data only. When applying the classifiers on other data sets (generalization of the model to $5$\um and $15$\um fibronectin), the features of each test condition are standardized with the mean and standard deviation of that condition. This removes differences in average cell morphology between conditions while using no information about the transition times of the test cells.\par

To ensure that the classifiers are always tested on cells unseen during training, we used repeated $K$-fold cross-validation grouped by cell. In each repetition $r=1,\dots,R$ the $29$ cells were randomly permuted and assigned to $K=4$ folds of $7$-$8$ cells. For each fold $k$ a classifier $f_{r,k}$ was trained on the observations of the other three folds and used to predict the observations of fold $k$. This yields an out-of-fold prediction
\begin{equation}
  \hat y_i = f_{r,k(c(i))}(\mathbf{x}_i)
\end{equation}
for every observation, made by a classifier that has never seen the corresponding cell. We performed $100$ repetitions with different random fold assignments and report the mean and standard deviation over repetitions.\par
For each repetition and split point, the out-of-fold predictions of all $N$ observations are pooled into a single $2\times2$ confusion matrix with entries $N_{ab}$, the number of observations with true class $a$ and predicted class $b$. With $p_{ab}=N_{ab}/N$ and the marginals $p_a=\sum_b p_{ab}$ and $q_b=\sum_a p_{ab}$, the mutual information between true and predicted class is
\begin{equation}
  I(y;\hat y) = \sum_{a,b} p_{ab}\,\log_2 \frac{p_{ab}}{p_a\,q_b} ,
\end{equation}
with terms for which $p_{ab}=0$ set to zero. \par

The mutual information is bounded by the entropy of the true labels, $I(y;\hat y)\le H(y)$, with
\begin{equation}
  H(y) = -p\log_2 p - (1-p)\log_2(1-p),
\end{equation}
where $p(t_c)$ is the fraction of \textit{near-transit} snapshots at split point $t_c$. Since $p$ increases monotonically with $t_c$ (from $0.06$ at $t_c=5$~minutes to $0.97$ at $t_c=115$~minutes, see \textbf{Fig.~\ref{fig:ml-balance}(a)}), $H(y)$ is maximal at $t_c=45$~minutes, where the two classes are balanced, and small at both ends of the sweep. A maximum of the mutual information as a function of $t_c$ can therefore partly reflect the class balance rather than the classification skill. To remove this dependence we additionally report the normalized mutual information
\begin{equation}
  U(t_c) = \frac{I(y;\hat y)}{H(y)} \in [0,1],
\end{equation}
such that $U=0$ for a classifier that carries no information about the true class and $U=1$ for a perfect classifier. For the out-of-distribution conditions, $H(y)$ is computed from the class balance of the respective test data. Because $H(y)$ is small at the ends of the sweep, $U$ is more sensitive to statistical fluctuations there.\par

We compared the performance of the classifiers with a null model in which the relation between features and time to transition is destroyed. Within each training set, the times to transition $T_i$ are randomly permuted among the training observations, and labels are assigned from the permuted times according to the same rule $T_i\le t_c$. This preserves the class balance of the training set at every $t_c$, while any association between $\mathbf{x}_i$ and $T_i$ is removed. The null classifier is trained on these permuted labels and evaluated on the unchanged held-out fold, and its mutual information is computed exactly as for the real classifier. The null is recomputed in each fold and repetition. \par
To test how robust the position of the maximum of the normalized mutual information $U$ is, we perform a bootstrap over trajectories. We resample the $29$ trajectories of the $15$\um data with replacement $1000$ times and compute $U$ for each $t_c$ in each resample from the out-of-fold predictions of the drawn cells (no retraining). We find that the maximum of $U$ is most often located at $t_c = 30$~minutes: in $64\%$ (SVC-RBF) and $49\%$ (LR) of the resamples, and at no other $t_c$ in more than $13\%$. Its position nevertheless varies considerably between resamples ($95\%$ interval $10-115$~minutes for SVC-RBF). For SVC-RBF, $U$ at $t_c = 30$~minutes is significantly higher than at its minimum ($t_c = 95$~minutes; $p = 0.022$, Bonferroni-corrected for $22$ comparisons), whereas for LR none of the comparisons with other $t_c$ is significant. The difference between $t_c = 25$-$35$~minutes and $t_c = 45$-$80$~minutes is not significant (SVC-RBF: $p = 0.16$; LR: $p = 0.38$), so the maximum is not sharply defined. Thus, the pre-transit signal is most likely strongest for split points around $30$~minutes, and it remains above the null model over most of the range of $t_c$.

\begin{table*}
\centering
\begin{tabular}{llcccccc}
\hline
Feature set & Classifier & $C=0.001$ & $C=0.01$ & $C=0.1$ & $C=1$ & $C=10$ & $C=100$ \\
\hline
\hline
\multicolumn{8}{l}{\textit{Mutual information [$10^{-3}$ bits]}} \\
best 2 & SVC-RBF & 0.9 & 5.0 & 26.7 & 25.6 & 27.4 & 24.7 \\
best 2 & LR & 18.2 & 18.4 & 18.4 & 18.4 & 18.4 & 18.4 \\
best 5 & SVC-RBF & 0.8 & 0.7 & 18.9 & 23.4 & 22.9 & 18.4 \\
best 5 & LR & 15.6 & 16.2 & 16.3 & 16.4 & 16.4 & 16.4 \\
best 7 & SVC-RBF & 0.8 & 0.7 & 20.0 & 15.4 & 16.9 & 11.3 \\
best 7 & LR & 13.3 & 13.7 & 14.3 & 15.7 & 16.4 & 16.5 \\
base & SVC-RBF & 0.8 & 0.6 & 11.7 & 12.5 & 16.0 & 13.0 \\
base & LR & 11.4 & 11.3 & 13.1 & 13.0 & 12.9 & 12.8 \\
all & SVC-RBF & 0.8 & 0.6 & 15.6 & 15.1 & 14.6 & 11.9 \\
all & LR & 13.0 & 15.5 & 18.7 & 20.3 & 21.3 & 21.5 \\
\hline
\multicolumn{8}{l}{\textit{PR-AUC}} \\
best 2 & SVC-RBF & 0.581 & 0.588 & 0.622 & 0.650 & 0.653 & 0.633 \\
best 2 & LR & 0.581 & 0.580 & 0.580 & 0.580 & 0.580 & 0.580 \\
best 5 & SVC-RBF & 0.612 & 0.605 & 0.628 & 0.639 & 0.623 & 0.615 \\
best 5 & LR & 0.632 & 0.631 & 0.629 & 0.629 & 0.629 & 0.629 \\
best 7 & SVC-RBF & 0.604 & 0.599 & 0.614 & 0.626 & 0.627 & 0.606 \\
best 7 & LR & 0.615 & 0.616 & 0.613 & 0.615 & 0.616 & 0.616 \\
base & SVC-RBF & 0.608 & 0.600 & 0.610 & 0.614 & 0.607 & 0.592 \\
base & LR & 0.595 & 0.598 & 0.606 & 0.610 & 0.610 & 0.610 \\
all & SVC-RBF & 0.608 & 0.601 & 0.620 & 0.630 & 0.619 & 0.601 \\
all & LR & 0.611 & 0.616 & 0.626 & 0.630 & 0.630 & 0.630 \\
\hline
\end{tabular}
\caption{\textbf{Sensitivity of the transit-time classifiers to the regularization parameter $C$.} Mutual information between predicted and true labels, and PR-AUC (near-transit as positive class), of SVC-RBF and LR, averaged over the split points $t_c = 20, 30, 45, 60, 80, 100$~minutes and 5 random divisions of the cells into 4 folds (cell-grouped cross-validation on $15$~\um data). SVC-RBF uses the median-heuristic $\gamma$ of each feature set. For $C \le 0.01$, the mutual information of SVC-RBF drops to nearly zero, suggesting its predicted labels carry almost no information about the true labels. For $C \in [0.1, 10]$, SVC-RBF performance is approximately stable. LR is largely insensitive to $C$, except for the \textit{all} feature set, whose performance increases up to $C \approx 1$. In the classification task for the prediction of transit times in the main text we use $C = 1$.}
\label{tab:C-sensitivity}
\end{table*}

\paragraph*{Data and code availability}
The data for the features and the python script used to calculate them are available on Zenodo (DOI: 10.5281/zenodo.23344526) and the scripts for the machine-learning analysis are available on GitHub \url{https://github.com/statbiophys/cell-transit}.

\section*{Acknowledgments}

This work was supported by an institutional Institut Curie grant (in 2020-2021) and by the French Agence Nationale de la Recherche (ANR), under grant ANR-22-CE30-0030 (project SQUEEZACTIV) and ANR-24-CE95-0005 (project DEMOQCELL). CS and MM thank Thierry Dubois for advice in cell handling.

\bibliographystyle{apsrev4-1}
\bibliography{bibliography}

\setcounter{figure}{0}
\renewcommand{\thefigure}{S\arabic{figure}}
\setcounter{table}{0}
\renewcommand{\thetable}{S\arabic{table}}

\newpage
\onecolumngrid

\section*{Supplementary Information}

\begin{figure}[h!]
    \includegraphics[width=\textwidth,
    trim={0 2cm 0 0},
    clip]{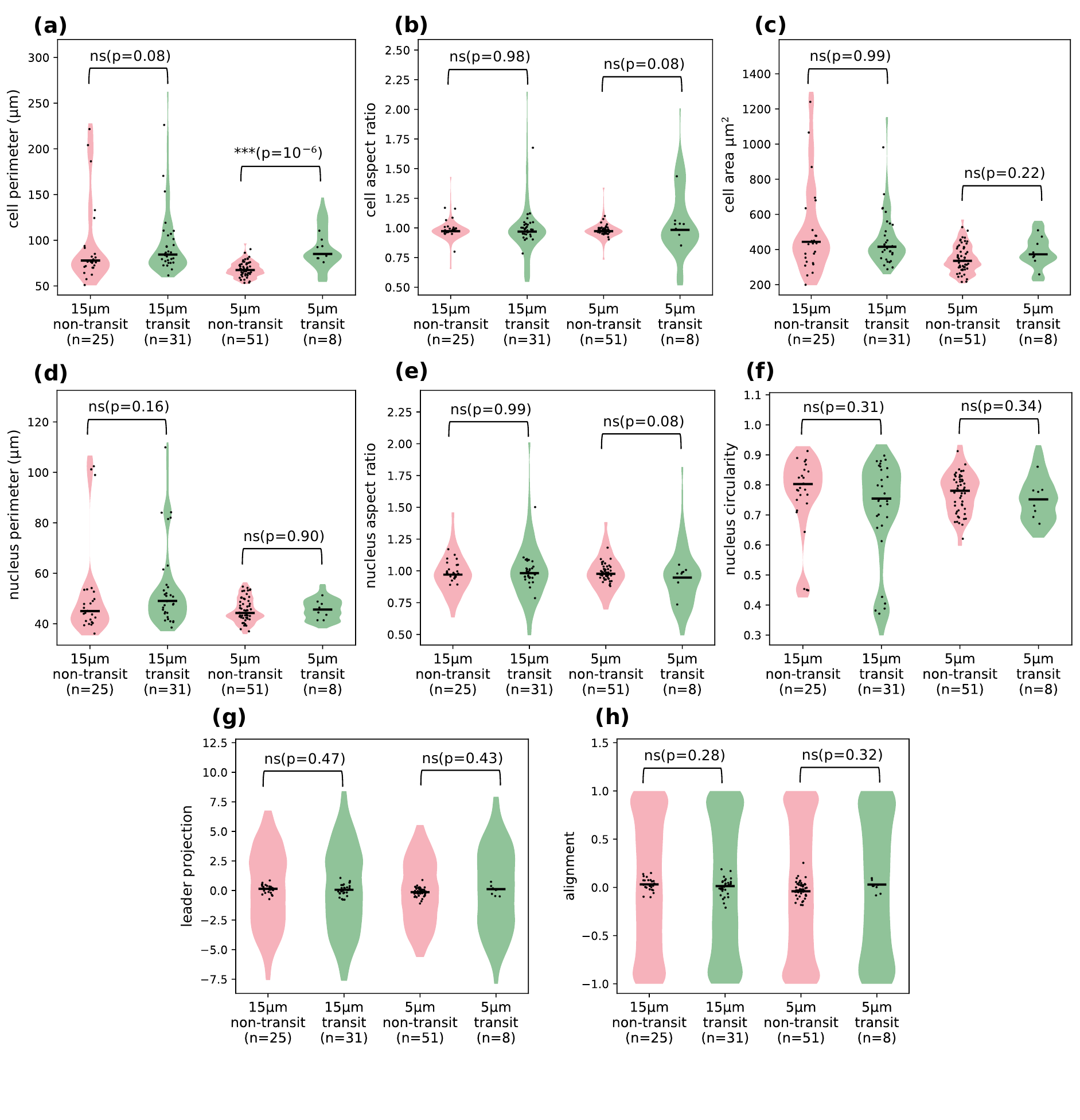}
    \caption{Distributions of contour-derived features for every snapshot of transiting (transit, red) and non-transiting (non-transit, green) cells in 15 and 5\um constriction patterns.\textbf{(a)} \textit{cell perimeter}, \textbf{(b)} \textit{cell aspect-ratio} \textbf{(c)} \textit{cell area} \textbf{(d)} \textit{nucleus perimeter}, \textbf{(e)} \textit{nucleus aspect-ratio}, \textbf{(f)} \textit{nucleus circularity}, \textbf{(g)} \textit{leader-projection}, and \textbf{(h)} \textit{alignment}. The red and green violin area represent data from snapshots at all timeframes, whereas the black points represent time-averaged values for individual trajectories. 
    The statistical significance is reported based on the $p$ value from the Mann-Whitney U test performed on time-averaged values to avoid cells tracked for longer times skewing the data.}
    \label{fig:SI_distributions}
\end{figure}

\begin{table*}[ht]
\centering
\footnotesize
\setlength{\tabcolsep}{4pt}
\caption{PR-AUC per test context: mean over the 100 splits [95\,\% CI], models tuned for PR-AUC. 15\,$\mu$m: corrected resampled $t$ over the splits (Nadeau \& Bengio 2003); 5\,$\mu$m and fibronectin: stratified cell bootstrap (2000 replicates, whole cells resampled within class, replicate $b$ scored with the model of split $b \bmod 100$). Chance: prevalence.}
\begin{tabular}{llccc}
\hline
Feature set & Model & 15\,$\mu$m & 5\,$\mu$m & Fibro 15\,$\mu$m \\
\hline
Best 2 & SVC-RBF & 0.653 [0.50, 0.81] & 0.702 [0.36, 0.93] & 0.701 [0.45, 0.90] \\
 & LR & 0.652 [0.50, 0.80] & 0.670 [0.31, 0.92] & 0.708 [0.47, 0.90] \\
Best 5 & SVC-RBF & 0.731 [0.56, 0.91] & 0.314 [0.10, 0.67] & 0.530 [0.35, 0.75] \\
 & LR & 0.529 [0.34, 0.72] & 0.226 [0.09, 0.53] & 0.562 [0.35, 0.75] \\
Best 7 & SVC-RBF & 0.733 [0.59, 0.88] & 0.577 [0.21, 0.90] & 0.618 [0.37, 0.83] \\
 & LR & 0.638 [0.46, 0.82] & 0.393 [0.15, 0.76] & 0.651 [0.38, 0.86] \\
Base Feat. & SVC-RBF & 0.673 [0.50, 0.84] & 0.619 [0.24, 0.90] & 0.593 [0.37, 0.84] \\
 & LR & 0.580 [0.39, 0.76] & 0.624 [0.28, 0.88] & 0.606 [0.34, 0.83] \\
All & SVC-RBF & 0.680 [0.50, 0.86] & 0.540 [0.17, 0.87] & 0.627 [0.40, 0.83] \\
 & LR & 0.632 [0.46, 0.81] & 0.327 [0.13, 0.67] & 0.639 [0.39, 0.84] \\
\hline
Chance &  & 0.290 & 0.022 & 0.415 \\
\hline
\end{tabular}
\label{tab:prauc-ci}
\end{table*}

\begin{table*}[ht]
\centering
\footnotesize
\setlength{\tabcolsep}{3pt}
\caption{Paired differences in PR-AUC, $\Delta = A - B$ (mean over the 100 splits) [95\,\% CI], models tuned for PR-AUC. 15\,$\mu$m: corrected resampled $t$ over the splits (Nadeau \& Bengio 2003); 5\,$\mu$m and fibronectin: stratified cell bootstrap (2000 replicates, whole cells resampled within class, replicate $b$ scored with the model of split $b \bmod 100$). Bold: the CI excludes 0. $^{*}$: Holm-adjusted $p < 0.05$ within the 25 comparisons of one context. CIs are not multiplicity-adjusted.}
\label{tab:diff_ci_ap}
\begin{tabular}{lccc}
\hline
Comparison ($A - B$) & 15\,$\mu$m & 5\,$\mu$m & Fibro 15\,$\mu$m \\
\hline
\multicolumn{4}{l}{\textit{SVC-RBF: feature sets}} \\
\quad Best 2 $-$ Best 5 & $-$0.08 [$-$0.28, 0.12] & \textbf{+0.39 [0.05, 0.62]} & +0.17 [$-$0.01, 0.33] \\
\quad Best 2 $-$ Best 7 & $-$0.08 [$-$0.23, 0.07] & +0.13 [$-$0.08, 0.43] & +0.08 [$-$0.08, 0.29] \\
\quad Best 2 $-$ Base Feat. & $-$0.02 [$-$0.17, 0.13] & +0.08 [$-$0.10, 0.38] & +0.11 [$-$0.06, 0.28] \\
\quad Best 2 $-$ All & $-$0.03 [$-$0.18, 0.13] & +0.16 [$-$0.07, 0.52] & +0.07 [$-$0.08, 0.24] \\
\quad Best 5 $-$ Best 7 & $-$0.00 [$-$0.16, 0.15] & $-$0.26 [$-$0.53, 0.06] & $-$0.09 [$-$0.25, 0.12] \\
\quad Best 5 $-$ Base Feat. & +0.06 [$-$0.09, 0.21] & $-$0.30 [$-$0.54, 0.02] & $-$0.06 [$-$0.22, 0.09] \\
\quad Best 5 $-$ All & +0.05 [$-$0.09, 0.19] & $-$0.23 [$-$0.50, 0.17] & $-$0.10 [$-$0.24, 0.06] \\
\quad Best 7 $-$ Base Feat. & +0.06 [$-$0.01, 0.13] & $-$0.04 [$-$0.28, 0.19] & +0.02 [$-$0.10, 0.14] \\
\quad Best 7 $-$ All & +0.05 [$-$0.04, 0.15] & +0.04 [$-$0.29, 0.45] & $-$0.01 [$-$0.15, 0.12] \\
\quad Base Feat. $-$ All & $-$0.01 [$-$0.09, 0.07] & +0.08 [$-$0.16, 0.44] & $-$0.03 [$-$0.11, 0.07] \\
\hline
\multicolumn{4}{l}{\textit{LR: feature sets}} \\
\quad Best 2 $-$ Best 5 & +0.12 [$-$0.08, 0.33] & \textbf{+0.44 [0.11, 0.68]} & +0.15 [$-$0.03, 0.36] \\
\quad Best 2 $-$ Best 7 & +0.01 [$-$0.14, 0.17] & +0.28 [$-$0.01, 0.51] & +0.06 [$-$0.11, 0.32] \\
\quad Best 2 $-$ Base Feat. & +0.07 [$-$0.06, 0.21] & +0.05 [$-$0.08, 0.22] & +0.10 [$-$0.06, 0.35] \\
\quad Best 2 $-$ All & +0.02 [$-$0.13, 0.17] & \textbf{+0.34 [0.05, 0.58]} & +0.07 [$-$0.09, 0.31] \\
\quad Best 5 $-$ Best 7 & $-$0.11 [$-$0.26, 0.04] & $-$0.17 [$-$0.42, 0.02] & $-$0.09 [$-$0.26, 0.12] \\
\quad Best 5 $-$ Base Feat. & $-$0.05 [$-$0.23, 0.13] & \textbf{$-$0.40 [$-$0.65, $-$0.07]} & $-$0.04 [$-$0.21, 0.16] \\
\quad Best 5 $-$ All & $-$0.10 [$-$0.26, 0.05] & $-$0.10 [$-$0.35, 0.08] & $-$0.08 [$-$0.22, 0.09] \\
\quad Best 7 $-$ Base Feat. & +0.06 [$-$0.03, 0.14] & $-$0.23 [$-$0.46, 0.03] & +0.05 [$-$0.09, 0.17] \\
\quad Best 7 $-$ All & +0.01 [$-$0.05, 0.06] & +0.07 [$-$0.04, 0.21] & +0.01 [$-$0.12, 0.11] \\
\quad Base Feat. $-$ All & $-$0.05 [$-$0.13, 0.02] & \textbf{+0.30 [0.04, 0.53]} & $-$0.03 [$-$0.15, 0.06] \\
\hline
\multicolumn{4}{l}{\textit{SVC-RBF $-$ LR, per feature set}} \\
\quad Best 2 & +0.00 [$-$0.11, 0.11] & +0.03 [$-$0.04, 0.11] & $-$0.01 [$-$0.10, 0.07] \\
\quad Best 5 & \textbf{+0.20 [0.03, 0.38]} & +0.09 [$-$0.15, 0.39] & $-$0.03 [$-$0.16, 0.13] \\
\quad Best 7 & +0.10 [$-$0.02, 0.21] & +0.18 [$-$0.19, 0.48] & $-$0.03 [$-$0.24, 0.22] \\
\quad Base Feat. & +0.09 [$-$0.03, 0.22] & $-$0.01 [$-$0.32, 0.22] & $-$0.01 [$-$0.18, 0.22] \\
\quad All & +0.05 [$-$0.03, 0.13] & +0.21 [$-$0.08, 0.47] & $-$0.01 [$-$0.17, 0.19] \\
\hline
\end{tabular}
\label{tab:prauc-ci-diff}
\end{table*}

\begin{figure}
    \centering
    \includegraphics[width=0.75\textwidth]{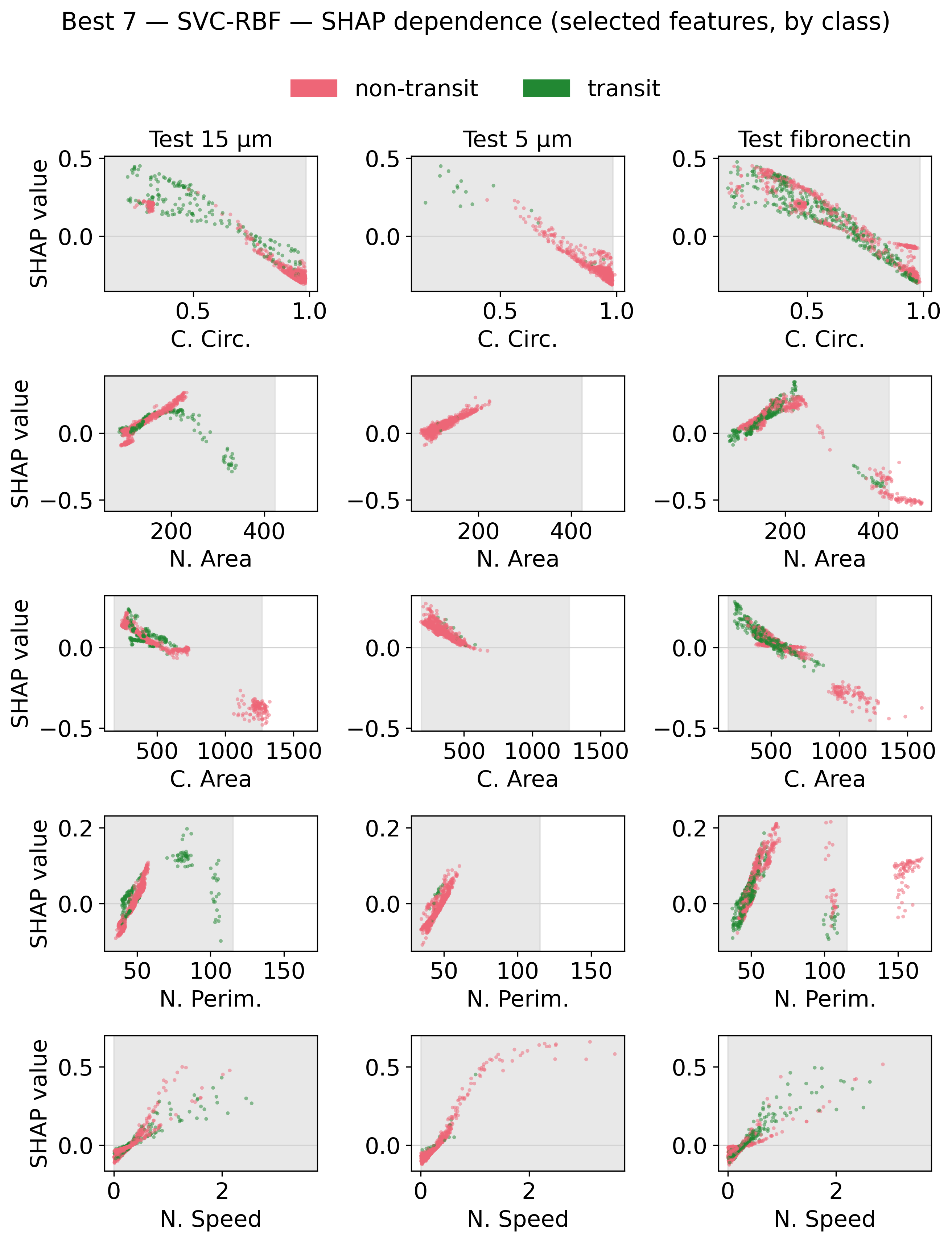}
    \caption{Shapley values of \textit{cell circularity, cell area, nucleus area, nucleus perimeter} and \textit{nucleus speed}. Shaded area shows the range of feature values in the training set of the $15$\um constriction. Green points corresponds to transiting trajectories and red to non-transiting. On the example of cell area and nucleus area we can see that the trajectories in the $5$\um constriction have a very limited values compared to $15$\um constriction with or without fibronectin.}
    \label{fig:shap-feature-comparison-15-5-fibro}
\end{figure}

\begin{figure}
    \centering
    \includegraphics[width=\textwidth]{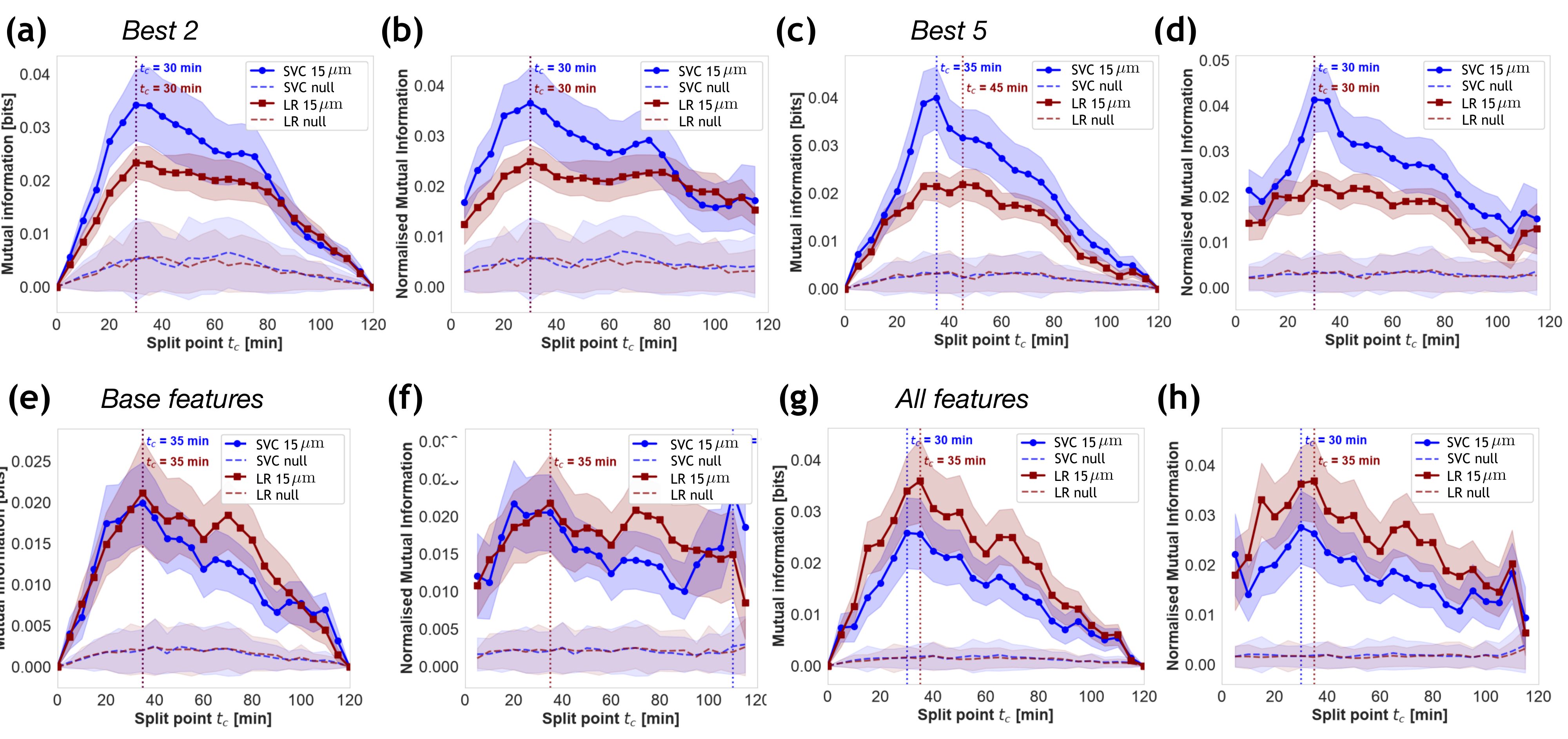}
    \caption{\textbf{Results for the prediction of transit times for \textit{best 2, best 5, base all} feature sets.}     
    Solid lines show the mean over the $100$ repetitions and shaded areas the standard deviation. We use Support Vector Classifier with radial basis function kernel (SVC-RBF, blue) and Logistic Regression (LR, red). Dashed lines show the null model, in which the transit times of the training time points were randomly permuted before labeling, preserving the class balance. 
    Classifiers trained and tested on $15$\um constrictions using the \textbf{(a-b)} \textit{best 2}, \textbf{(c-d)} \textit{best 5}, \textbf{(e-f)} \textit{base} and \textbf{(g-h)} \textit{all} feature set. Mutual information $I$ between true and predicted class as a function of the split point $t_c$ as well as normalized mutual information. The maximum of both classifiers is marked by a dotted line and it stays at around $t_c=30-35$~minutes for all feature sets. Both classifiers perform above chance at all split points.}
    \label{fig:res-all}
\end{figure}

\begin{figure}
    \centering
    \includegraphics[width=0.33\textwidth]{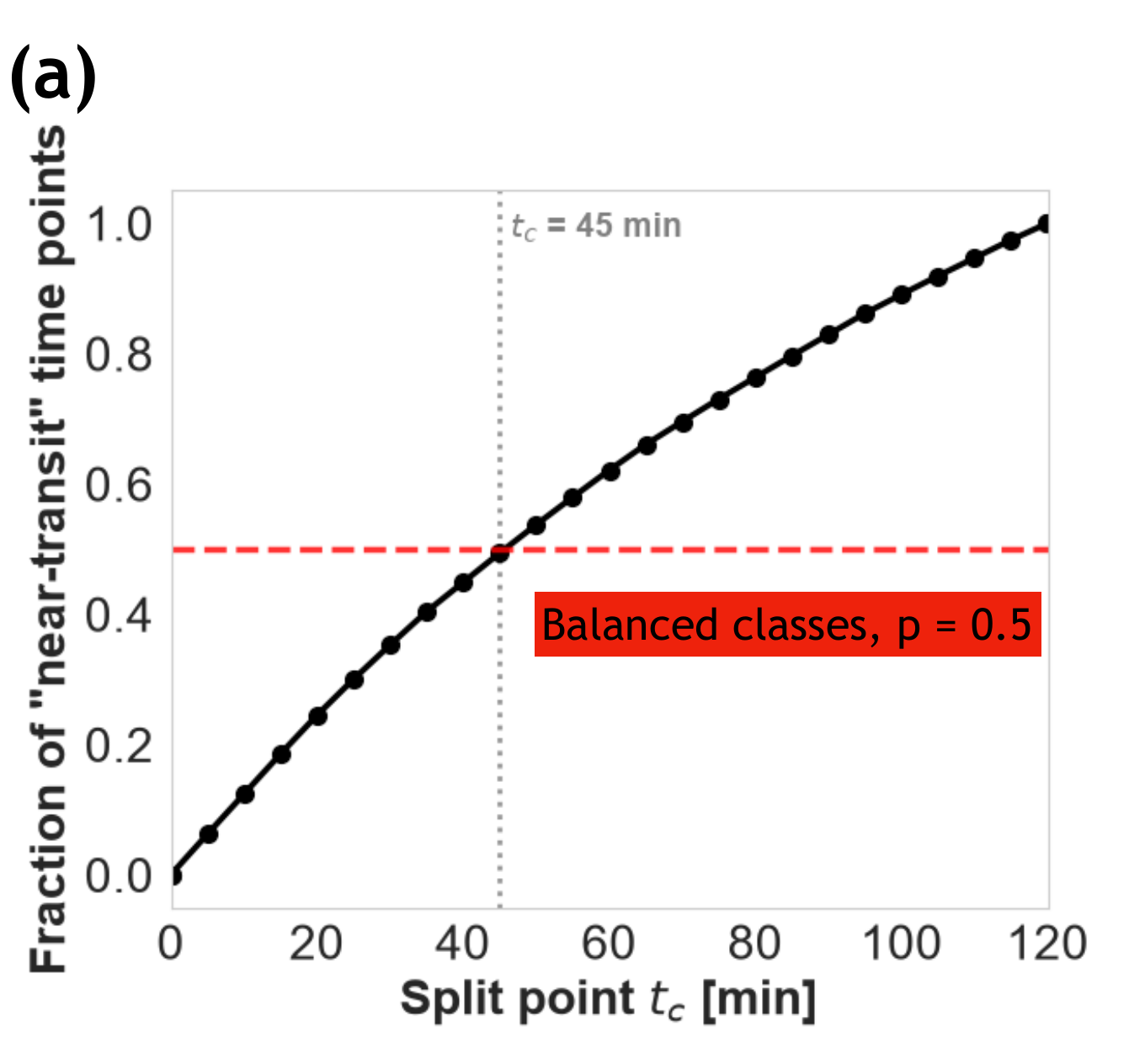} 
    \caption{\textbf{Fraction of \textit{near-transit} trajectories in the dataset} \textbf{(a)} Fraction $p(t_c)$ of time points labeled as \textit{near-transit} as a function of the split point $t_c$ ($15$\um constrictions data). Since every time point with a time to transit $\Delta t\le t_c$ is labeled \textit{near-transit}, $p(t_c)$ is the cumulative distribution of the times to transit. The two classes are balanced ($p=0.5$, red dashed line) at $t_c=45$~minutes.}
    \label{fig:ml-balance}
\end{figure}

\end{document}

%% file: preamble.tex
\usepackage{upgreek}
\usepackage{graphicx}
\usepackage{upgreek}
\usepackage{graphicx}
\usepackage{overpic}
\usepackage{dirtytalk}
\usepackage[hidelinks]{hyperref}
\usepackage{mathtools}
\usepackage{amssymb}
\usepackage{xspace}
\newcommand{\um}{$\upmu\mathrm{m}$\xspace}
\newcommand{\ums}{$\upmu\mathrm{m}^2$\xspace}
\usepackage{hyperref}
\usepackage[nameinlink,noabbrev]{cleveref}
\usepackage{multirow}
\usepackage{bm}
\usepackage{xcolor}
\usepackage[hidelinks]{hyperref}   
\usepackage{upgreek}
\usepackage{graphicx}
\usepackage{upgreek}
\usepackage{graphicx}
\usepackage{overpic}
\usepackage{lipsum}
\usepackage{booktabs}
\usepackage{booktabs}
\usepackage{comment}
\usepackage{xcolor} 

\definecolor{darkgreen}{rgb}{0.0, 0.5, 0.0}

\usepackage{subfiles}
\usepackage{comment}
\usepackage{booktabs}
\usepackage{xcolor} 
\definecolor{darkgreen}{rgb}{0.0, 0.5, 0.0}

\renewcommand{\thetable}{\arabic{table}}

%% file: bibliography.bib
@article{hirway2023review,
  title={A review of computational modeling, machine learning and image analysis in cancer metastasis dynamics},
  author={Hirway, Shreyas U and Weinberg, Seth H},
  journal={Computational and Systems Oncology},
  volume={3},
  number={1},
  pages={e1044},
  year={2023},
  publisher={Wiley Online Library}
}

@article{zhang2018morphology,
  title={Morphology-based prediction of cancer cell migration using an artificial neural network and a random decision forest},
  author={Zhang, Zhixiong and Chen, Lili and Humphries, Brock and Brien, Riley and Wicha, Max S and Luker, Kathryn E and Luker, Gary D and Chen, Yu-Chih and Yoon, Euisik},
  journal={Integrative Biology},
  volume={10},
  number={12},
  pages={758--767},
  year={2018},
  publisher={Oxford University Press}
}

@article{eddy2021morphodynamics,
  title={Morphodynamics facilitate cancer cells to navigate 3D extracellular matrix},
  author={Eddy, Christopher Z and Raposo, Helena and Manchanda, Aayushi and Wong, Ryan and Li, Fuxin and Sun, Bo},
  journal={Scientific reports},
  volume={11},
  number={1},
  pages={20434},
  year={2021},
  publisher={Nature Publishing Group UK London}
}

@article{kimmel2019deep,
  title={Deep convolutional and recurrent neural networks for cell motility discrimination and prediction},
  author={Kimmel, Jacob C and Brack, Andrew S and Marshall, Wallace F},
  journal={IEEE/ACM transactions on Computational Biology and Bioinformatics},
  volume={18},
  number={2},
  pages={562--574},
  year={2019},
  publisher={IEEE}
}

@article{mencattini2020discovering,
  title={Discovering the hidden messages within cell trajectories using a deep learning approach for in vitro evaluation of cancer drug treatments},
  author={Mencattini, A and Di Giuseppe, D and Comes, MC and Casti, P and Corsi, F and Bertani, FR and Ghibelli, L and Businaro, L and Di Natale, C and Parrini, MC and others},
  journal={Scientific reports},
  volume={10},
  number={1},
  pages={7653},
  year={2020},
  publisher={Nature Publishing Group UK London}
}

@article{bergert2012cell,
  title={Cell mechanics control rapid transitions between blebs and lamellipodia during migration},
  author={Bergert, Martin and Chandradoss, Stanley D and Desai, Ravi A and Paluch, Ewa},
  journal={Proceedings of the National Academy of Sciences},
  volume={109},
  number={36},
  pages={14434--14439},
  year={2012},
  publisher={National Academy of Sciences}
}

@article{bahar2021modeling,
  title={Modeling and predicting the cell migration properties from scratch wound healing assay on cisplatin-resistant ovarian cancer cell lines using artificial neural network},
  author={Bahar, Entaz and Yoon, Hyonok},
  booktitle={Healthcare},
  volume={9},
  number={7},
  pages={911},
  year={2021},
  organization={MDPI}
}

@article{umar2023prediction,
  title={Prediction of cell migration potential on human breast cancer cells treated with Albizia lebbeck ethanolic extract using extreme machine learning},
  author={Umar, Huzaifa and Aliyu, Maryam Rabiu and Usman, Abdullahi Garba and Ghali, Umar Muhammad and Abba, Sani Isah and Ozsahin, Dilber Uzun},
  journal={Scientific Reports},
  volume={13},
  number={1},
  pages={22242},
  year={2023},
  publisher={Nature Publishing Group UK London}
}

@article{van2025deformability,
  title={Deformability determines confined cancer cell migration efficiency with limited effect on directionality},
  author={van der Net, A and Boot, RC and Van Dijk, I and Conboy, JP and Boukany, PE and Koenderink, GH},
  journal={Biomicrofluidics},
  volume={19},
  number={6},
  year={2025},
  publisher={AIP Publishing}
}

@article{ruprecht2015cortical,
  title={Cortical contractility triggers a stochastic switch to fast amoeboid cell motility},
  author={Ruprecht, Verena and Wieser, Stefan and Callan-Jones, Andrew and Smutny, Michael and Morita, Hitoshi and Sako, Keisuke and Barone, Vanessa and Ritsch-Marte, Monika and Sixt, Michael and Voituriez, Rapha{\"e}l and others},
  journal={Cell},
  volume={160},
  number={4},
  pages={673--685},
  year={2015},
  publisher={Elsevier}
}

@article{arjonen2011filopodia,
  title={Filopodia and adhesion in cancer cell motility},
  author={Arjonen, Antti and Kaukonen, Riina and Ivaska, Johanna},
  journal={Cell adhesion \& migration},
  volume={5},
  number={5},
  pages={421--430},
  year={2011},
  publisher={Taylor \& Francis}
}

@article{roche2018epithelial,
  title={The epithelial-to-mesenchymal transition in cancer},
  author={Roche, Jo{\"e}lle},
  journal={Cancers},
  volume={10},
  number={2},
  pages={52},
  year={2018},
  publisher={MDPI}
}

@article{karling2025immune,
  title={Immune cells adapt to confined environments in vivo to optimise nuclear plasticity for migration},
  author={Karling, Tua and Weavers, Helen},
  journal={EMBO reports},
  volume={26},
  number={5},
  pages={1238},
  year={2025}
}

@article{amiri2024inferring,
  title={Inferring geometrical dynamics of cell nucleus translocation},
  author={Amiri, Sirine and Zhang, Yirui and Gerardos, Andonis and Sykes, C{\'e}cile and Ronceray, Pierre},
  journal={Physical Review Research},
  volume={6},
  number={4},
  pages={043030},
  year={2024},
  publisher={APS}
}

@article{amiri2025nuclear,
  title={The nuclear lamin network passively responds to both active or passive cell movement through confinements},
  author={Amiri, Sirine and Bos, Inge and Reyssat, Etienne and Sykes, C{\'e}cile},
  journal={Soft matter},
  volume={21},
  number={5},
  pages={893--902},
  year={2025},
  publisher={The Royal Society of Chemistry}
}

@article{bruckner2022geometry,
  title={Geometry adaptation of protrusion and polarity dynamics in confined cell migration},
  author={Br{\"u}ckner, David B and Schmitt, Matthew and Fink, Alexandra and Ladurner, Georg and Flommersfeld, Johannes and Arlt, Nicolas and Hannezo, Edouard and R{\"a}dler, Joachim O and Broedersz, Chase P},
  journal={Physical Review X},
  volume={12},
  number={3},
  pages={031041},
  year={2022},
  publisher={APS}
}

@article{godeau20203d,
  title={3D single cell migration driven by temporal correlation between oscillating force dipoles},
  author={Godeau, Amelie Luise and Leoni, Marco and Comelles, Jordi and Guyomar, Tristan and Lieb, Michele and Delanoe-Ayari, Helene and Ott, Albrecht and Harlepp, Sebastien and Sens, Pierre and Riveline, Daniel},
  journal={bioRxiv},
  pages={2020--05},
  year={2020},
  publisher={Cold Spring Harbor Laboratory}
}

@article{stoberl2024nuclear,
  title={Nuclear deformation and dynamics of migrating cells in 3D confinement reveal adaptation of pulling and pushing forces},
  author={St{\"o}berl, Stefan and Flommersfeld, Johannes and Kreft, Maximilian M and Benoit, Martin and Broedersz, Chase P and R{\"a}dler, Joachim O},
  journal={Science advances},
  volume={10},
  number={34},
  pages={eadm9195},
  year={2024},
  publisher={American Association for the Advancement of Science}
}

@article{venturini2020nucleus,
  title={The nucleus measures shape changes for cellular proprioception to control dynamic cell behavior},
  author={Venturini, Valeria and Pezzano, Fabio and Catala Castro, Frederic and H{\"a}kkinen, Hanna-Maria and Jim{\'e}nez-Delgado, Senda and Colomer-Rosell, Mariona and Marro, Monica and Tolosa-Ramon, Queralt and Paz-L{\'o}pez, Sonia and Valverde, Miguel A and others},
  journal={Science},
  volume={370},
  number={6514},
  pages={eaba2644},
  year={2020},
  publisher={American Association for the Advancement of Science}
}

@article{renkawitz2019nuclear,
  title={Nuclear positioning facilitates amoeboid migration along the path of least resistance},
  author={Renkawitz, J{\"o}rg and Kopf, Aglaja and Stopp, Julian and De Vries, Ingrid and Driscoll, Meghan K and Merrin, Jack and Hauschild, Robert and Welf, Erik S and Danuser, Gaudenz and Fiolka, Reto and others},
  journal={Nature},
  volume={568},
  number={7753},
  pages={546--550},
  year={2019},
  publisher={Nature Publishing Group UK London}
}

@article{lautscham2015migration,
  title={Migration in confined 3D environments is determined by a combination of adhesiveness, nuclear volume, contractility, and cell stiffness},
  author={Lautscham, Lena A and K{\"a}mmerer, Christoph and Lange, Janina R and Kolb, Thorsten and Mark, Christoph and Schilling, Achim and Strissel, Pamela L and Strick, Reiner and Gluth, Caroline and Rowat, Amy C and others},
  journal={Biophysical journal},
  volume={109},
  number={5},
  pages={900--913},
  year={2015},
  publisher={Elsevier}
}

@article{sadjadi2022ameboid,
  title={Ameboid cell migration through regular arrays of micropillars under confinement},
  author={Sadjadi, Zeinab and Vesperini, Doriane and Laurent, Annalena M and Barnefske, Lena and Terriac, Emmanuel and Lautenschl{\"a}ger, Franziska and Rieger, Heiko},
  journal={Biophysical Journal},
  volume={121},
  number={23},
  pages={4615--4623},
  year={2022},
  publisher={Elsevier}
}

@article{scikit-learn,
  title={Scikit-learn: Machine Learning in {P}ython},
  author={Pedregosa, F. and Varoquaux, G. and Gramfort, A. and Michel, V.
          and Thirion, B. and Grisel, O. and Blondel, M. and Prettenhofer, P.
          and Weiss, R. and Dubourg, V. and Vanderplas, J. and Passos, A. and
          Cournapeau, D. and Brucher, M. and Perrot, M. and Duchesnay, E.},
  journal={Journal of Machine Learning Research},
  volume={12},
  pages={2825--2830},
  year={2011}
}

@incollection{NIPS2017_7062,
title = {A Unified Approach to Interpreting Model Predictions},
author = {Lundberg, Scott M and Lee, Su-In},
year = {2017},
publisher = {Curran Associates, Inc.},
url = {http://papers.nips.cc/paper/7062-a-unified-approach-to-interpreting-model-predictions.pdf}
}

@inbook{Shapley1953,
  title = {17. A Value for n-Person Games},
  ISBN = {9781400881970},
  url = {http://dx.doi.org/10.1515/9781400881970-018},
  DOI = {10.1515/9781400881970-018},
  booktitle = {Contributions to the Theory of Games (AM-28),  Volume II},
  publisher = {Princeton University Press},
  author = {Shapley,  L. S.},
  year = {1953},
  month = Dec,
  pages = {307–318}
}

@article{paul2017cancer,
  title={Cancer cell motility: lessons from migration in confined spaces},
  author={Paul, Colin D and Mistriotis, Panagiotis and Konstantopoulos, Konstantinos},
  journal={Nature reviews cancer},
  volume={17},
  number={2},
  pages={131--140},
  year={2017},
  publisher={Nature Publishing Group UK London}
}

@misc{yolov8,
  author = {{Ultralytics}},
  title  = {Ultralytics YOLOv8},
  year   = {2023},
  url    = {https://github.com/ultralytics/ultralytics},
  note   = {Accessed: 2024-10-23}
}

@article{perea2021squeezing,
  title={Squeezing through the microcirculation: survival adaptations of circulating tumour cells to seed metastasis: Metastasis},
  author={Perea Paizal, Julia and Au, Sam H and Bakal, Chris},
  journal={British Journal of Cancer},
  volume={124},
  number={1},
  pages={58--65},
  year={2021},
  publisher={Nature Publishing Group UK London}
}

@article{salminen2020vitro,
  title={In vitro studies of transendothelial migration for biological and drug discovery},
  author={Salminen, Alec T and Allahyari, Zahra and Gholizadeh, Shayan and McCloskey, Molly C and Ajalik, Raquel and Cottle, Renee N and Gaborski, Thomas R and McGrath, James L},
  journal={Frontiers in medical technology},
  volume={2},
  pages={600616},
  year={2020},
  publisher={Frontiers Media SA}
}

@article{mcgregor2016squish,
  title={Squish and squeeze—the nucleus as a physical barrier during migration in confined environments},
  author={McGregor, Alexandra Lynn and Hsia, Chieh-Ren and Lammerding, Jan},
  journal={Current opinion in cell biology},
  volume={40},
  pages={32--40},
  year={2016},
  publisher={Elsevier}
}

@article{wortel2021local,
  title={Local actin dynamics couple speed and persistence in a cellular Potts model of cell migration},
  author={Wortel, Inge MN and Niculescu, Ioana and Kolijn, P Martijn and Gov, Nir S and de Boer, Rob J and Textor, Johannes},
  journal={Biophysical journal},
  volume={120},
  number={13},
  pages={2609--2622},
  year={2021},
  publisher={Elsevier}
}

@article{moure2017phase,
  title={Phase-field model of cellular migration: Three-dimensional simulations in fibrous networks},
  author={Moure, Adrian and Gomez, Hector},
  journal={Computer Methods in Applied Mechanics and Engineering},
  volume={320},
  pages={162--197},
  year={2017},
  publisher={Elsevier}
}

@article{gabbireddy2021melanoma,
  title={Melanoma cells adopt features of both mesenchymal and amoeboid migration within confining channels},
  author={Gabbireddy, Sairisheel R and Vosatka, Karl W and Chung, Aram J and Logue, Jeremy S},
  journal={Scientific reports},
  volume={11},
  number={1},
  pages={17804},
  year={2021},
  publisher={Nature Publishing Group UK London}
}

@article{holle2019cancer,
  title={Cancer cells invade confined microchannels via a self-directed mesenchymal-to-amoeboid transition},
  author={Holle, Andrew W and Govindan Kutty Devi, Neethu and Clar, Kim and Fan, Anthony and Saif, Taher and Kemkemer, Ralf and Spatz, Joachim P},
  journal={Nano letters},
  volume={19},
  number={4},
  pages={2280--2290},
  year={2019},
  publisher={ACS Publications}
}

@article{vaidvziulyte2022persistent,
  title={Persistent cell migration emerges from a coupling between protrusion dynamics and polarized trafficking},
  author={Vaid{\v{z}}iulyt{\.e}, Kotryna and Mac{\'e}, Anne-Sophie and Battistella, Aude and Beng, William and Schauer, Kristine and Coppey, Mathieu},
  journal={Elife},
  volume={11},
  pages={e69229},
  year={2022},
  publisher={eLife Sciences Publications, Ltd}
}

@article{todorovski2023confined,
  title={Confined environments induce polarized paraspeckle condensates},
  author={Todorovski, Vanja and McCluggage, Finn and Li, Yixuan and Meid, Annika and Spatz, Joachim P and Holle, Andrew W and Fox, Archa H and Choi, Yu Suk},
  journal={Communications Biology},
  volume={6},
  number={1},
  pages={145},
  year={2023},
  publisher={Nature Publishing Group UK London}
}

@article{tran2025cytoplasmic,
  title={Cytoplasmic anillin and Ect2 promote RhoA/myosin II-dependent confined migration and invasion},
  author={Tran, Avery T and Wisniewski, Emily O and Mistriotis, Panagiotis and Stoletov, Konstantin and Parlani, Maria and Amitrano, Alice and Ifemembi, Brent and Lee, Se Jong and Bera, Kaustav and Zhang, Yuqi and others},
  journal={Nature Materials},
  volume={24},
  number={9},
  pages={1476--1488},
  year={2025},
  publisher={Nature Publishing Group UK London}
}

@article{hsia2022confined,
  title={Confined migration induces heterochromatin formation and alters chromatin accessibility},
  author={Hsia, Chieh-Ren and McAllister, Jawuanna and Hasan, Ovais and Judd, Julius and Lee, Seoyeon and Agrawal, Richa and Chang, Chao-Yuan and Soloway, Paul and Lammerding, Jan},
  journal={Iscience},
  volume={25},
  number={9},
  year={2022},
  publisher={Elsevier}
}

@article{yamao2015distinct,
  title={Distinct predictive performance of Rac1 and Cdc42 in cell migration},
  author={Yamao, Masataka and Naoki, Honda and Kunida, Katsuyuki and Aoki, Kazuhiro and Matsuda, Michiyuki and Ishii, Shin},
  journal={Scientific reports},
  volume={5},
  number={1},
  pages={17527},
  year={2015},
  publisher={Nature Publishing Group UK London}
}

@article{mittal2018epithelial,
  title={Epithelial mesenchymal transition in tumor metastasis},
  author={Mittal, Vivek},
  journal={Annual Review of Pathology: Mechanisms of Disease},
  volume={13},
  pages={395--412},
  year={2018},
  publisher={Annual Reviews}
}

@article{charras2014physical,
  title={Physical influences of the extracellular environment on cell migration},
  author={Charras, Guillaume and Sahai, Erik},
  journal={Nature reviews Molecular cell biology},
  volume={15},
  number={12},
  pages={813--824},
  year={2014},
  publisher={Nature Publishing Group UK London}
}

@article{guck2010critical,
  title={Critical review: cellular mechanobiology and amoeboid migration},
  author={Guck, Jochen and Lautenschl{\"a}ger, Franziska and Paschke, Stephan and Beil, Michael},
  journal={Integrative biology},
  volume={2},
  number={11-12},
  pages={575--583},
  year={2010},
  publisher={The Royal Society of Chemistry}
}

@article{lammermann2009mechanical,
  title={Mechanical modes of ‘amoeboid’cell migration},
  author={L{\"a}mmermann, Tim and Sixt, Michael},
  journal={Current opinion in cell biology},
  volume={21},
  number={5},
  pages={636--644},
  year={2009},
  publisher={Elsevier}
}

@article{friedl2011nuclear,
  title={Nuclear mechanics during cell migration},
  author={Friedl, Peter and Wolf, Katarina and Lammerding, Jan},
  journal={Current opinion in cell biology},
  volume={23},
  number={1},
  pages={55--64},
  year={2011},
  publisher={Elsevier}
}

@article{keys2024rear,
  title={Rear cortex contraction aids in nuclear transit during confined migration by increasing pressure in the cell posterior},
  author={Keys, Jeremy and Cheung, Brian CH and Elpers, Margaret A and Wu, Mingming and Lammerding, Jan},
  journal={Journal of Cell Science},
  volume={137},
  number={12},
  pages={jcs260623},
  year={2024},
  publisher={The Company of Biologists Ltd}
}

@article{radler2026dynamic,
  title={Dynamic protrusions mediate crawling motility in Asgard archaea},
  author={Radler, Philipp and Viehboeck, Tobias and Luo, Zhen-Hao and Masla{\'c}, Nevena and Schmidt, Katharina and Hauschild, Robert and Nobu, Masaru K and Bulgheresi, Silvia and Stradal, Theresia EB and Rottner, Klemens and others},
  journal={Nature},
  pages={1--8},
  year={2026},
  publisher={Nature Publishing Group UK London}
}

@article{lin2019fibronectin,
  title={Fibronectin in cancer: friend or foe},
  author={Lin, Tsung-Cheng and Yang, Cheng-Han and Cheng, Li-Hsin and Chang, Wen-Tsan and Lin, Yuh-Rong and Cheng, Hung-Chi},
  journal={Cells},
  volume={9},
  number={1},
  pages={27},
  year={2019},
  publisher={MDPI}
}

@article{lugo2020breast,
  title={Breast fibroblasts and ECM components modulate breast cancer cell migration through the secretion of MMPs in a 3D microfluidic co-culture model},
  author={Lugo-Cintron, Karina M and Gong, Max M and Ayuso, Jose M and Tomko, Lucas A and Beebe, David J and Virumbrales-Munoz, Maria and Ponik, Suzanne M},
  journal={Cancers},
  volume={12},
  number={5},
  pages={1173},
  year={2020},
  publisher={MDPI}
}

@article{longstreth2024role,
  title={The role of fibronectin in mediating cell migration},
  author={Longstreth, Jessica H and Wang, Karin},
  journal={American Journal of Physiology-Cell Physiology},
  volume={326},
  number={4},
  pages={C1212--C1225},
  year={2024},
  publisher={American Physiological Society Rockville, MD}
}

@article{davidson2014nuclear,
  title={Nuclear deformability constitutes a rate-limiting step during cell migration in 3-D environments},
  author={Davidson, Patricia M and Denais, Celine and Bakshi, Maya C and Lammerding, Jan},
  journal={Cellular and molecular bioengineering},
  volume={7},
  number={3},
  pages={293--306},
  year={2014},
  publisher={Springer}
}
